\documentclass[english,prb,twocolumn,superscriptaddress,floatfix]{revtex4}
\usepackage{graphicx}
\usepackage{amssymb}
\usepackage{babel}
\usepackage{bm}

\newcommand{\ket}[1]{\ensuremath{\left| #1 \right\rangle }}
\newcommand{\bra}[1]{\ensuremath{\left\langle #1 \right|}}

\newcommand{\Jz}{ \left\langle J_z \right\rangle }
\newcommand{\Jx}{ \left\langle J_x \right\rangle }

\usepackage{color}  
\usepackage{ulem}

\graphicspath{{../}}

\begin{document}
\normalem

\title{Assessment of the RE(OH)$_{\bf 3}$ Ising-like Magnetic Materials as Possible
Candidates for the Study of Transverse-Field-Induced Quantum Phase Transitions}

\author{Pawel Stasiak}
\affiliation{Department of Physics and Astronomy, University of Waterloo, Waterloo, Ontario, N2L-3G1 Canada}

\author{Michel J.P. Gingras}
\affiliation{Department of Physics and Astronomy, University of Waterloo, Waterloo, Ontario, N2L-3G1 Canada}
\affiliation{Department of Physics and Astronomy, University of Canterbury, Private Bag 4800, Christchurch, New Zealand}
\affiliation{Canadian Institute for Advanced Research, 180 Dundas St. W., Toronto, Ontario, M5G 1Z8, Canada}
\date{\today}

\begin{abstract}
The LiHo$_x$Y$_{1-x}$F$_4$ Ising magnetic material subject to a magnetic field, $B_x$,
perpendicular to the Ho$^{3+}$ Ising direction has shown over the past twenty years
to be a host of very interesting thermodynamic and magnetic phenomena. 
Unfortunately, the availability of other magnetic materials other than
LiHo$_x$Y$_{1-x}$F$_4$ that may be described by a transverse field Ising model
remains very much limited. It is in this context that we use here mean-field theory
to investigate the suitability of the Ho(OH)$_{3}$, Dy(OH)$_{3}$ and Tb(OH)$_3$
insulating hexagonal dipolar Ising-like ferromagnets for the study of the quantum
phase transition induced by a magnetic field, $B_x$, applied perpendicular to the 
Ising spin direction.
Experimentally, the zero field critical (Curie) temperatures are known to be 
$T_{c} \approx 2.54$ K,
$T_{c} \approx 3.48$ K and 
$T_{c} \approx 3.72$ K, for Ho(OH)$_{3}$, Dy(OH)$_{3}$
and Tb(OH)$_{3}$, respectively. 
From our calculations we estimate
the critical transverse field, $B_{x}^{c}$, to destroy ferromagnetic
order at zero temperature to be 
$B_{x}^{c}=$4.35 T, 
$B_{x}^{c}=$5.03 T
and 
$B_{x}^{c}=$54.81 T 
for 
Ho(OH)$_{3}$, 
Dy(OH)$_{3}$ 
and Tb(OH)$_{3}$, respectively. 
We find that Ho(OH)$_{3}$, similarly to LiHoF$_4$, can be quantitatively 
described by an effective $S=1/2$ transverse field Ising model (TFIM). 
This is not the case for Dy(OH)$_{3}$ due to the strong admixing between
the ground doublet and first excited doublet induced by the dipolar
interactions. 
Furthermore, we find that the paramagnetic (PM) to ferromagnetic (FM) 
transition
in Dy(OH)$_3$ becomes first order for strong $B_x$ and low temperatures.
Hence, the PM to FM zero temperature transition in 
Dy(OH)$_{3}$ may be first order and not quantum critical.
We investigate the effect of competing antiferromagnetic nearest-neighbor exchange 
and applied magnetic field $B_z$ along the Ising spin direction $\hat{z}$ on the
first order transition in Dy(OH)$_{3}$.
We conclude from these preliminary calculations that Ho(OH)$_{3}$
and Dy(OH)$_{3}$, and their Y$^{3+}$ diamagnetically diluted variants,
Ho$_{x}$Y$_{1-x}$(OH)$_{3}$ and Dy$_{x}$Y$_{1-x}$(OH)$_{3}$,
are potentially interesting systems to study transverse-field induced
quantum fluctuations effects in hard axis (Ising-like) magnetic materials. 
\end{abstract}

\maketitle

\section{Introduction }

\subsection{Transverse field Ising model}

Quantum phase transitions occur near zero temperature and are driven
by quantum mechanical fluctuations associated with the Heisenberg
uncertainty principle and not by thermal fluctuations as in the case
of classical temperature-driven phase transitions~\cite{Sondhi,Sachdev}. 
There is accumulating evidence that the exotic behavior exhibited by 
several metallic, magnetic
and superconducting materials may have its origin in underlying large
quantum fluctuations and proximity to a quantum phase transition.
For this reason, much efforts are currently being devoted to understand
quantum phase transitions in a wide variety of condensed matter systems.

Perhaps the simplest model that embodies the phenomenon of a quantum
phase transition is the transverse field Ising model (TFIM) ~\cite{Pfeuty,Sen}, 
first proposed by de Gennes to describe proton tunneling in ferrolectric
materials~\cite{deGennes}. The spin Hamiltonian for the TFIM reads:
\begin{eqnarray}
H_{{\rm TFIM}} & = & -\frac{1}{2}\sum_{({\bm r}_{i},
{\bm r}_{j})}J_{ij}(\vert{\bm r}_{j}-{\bm r}_{i})\vert)S_{i}^{z}
({\bm r}_{i})S_{j}^{z}({\bm r}_{j})\label{TFIM}\\
 &  & -\Gamma\sum_{({\bm r}_{i})}S_{i}^{x}({\bm r}_{i}).
\nonumber 
\label{Htfim}
\end{eqnarray}
 The ${\bm S}_{i}$ ($S=1/2$) quantum spin operators reside on the
lattice sites ${\bm r}_{i}$ of some $d-$dimensional lattice. 
The ${\bm S}_i$ operators are related
to the Pauli spin matrices by ${\bm S}_i=\frac{1}{2} {\bm \sigma}_i$
(here we set $\hbar=1$).
The components of ${\bm S}_{i}$ obey the commutation relations 
$[S_{i}^{\alpha},S_{j}^{\beta}]=i\epsilon_{\alpha\beta\gamma}S_{i}^{\gamma}\delta_{ij}$
where $\alpha$, $\beta$ and $\gamma$ indicate $x$, $y$ and $z$
spin components, $\delta_{ij}$ and $\epsilon_{\alpha\beta\gamma}$
are the Kronecker delta and fully antisymmetric tensor, respectively. 
$\Gamma$ is the effective transverse field
along the $x$ direction, perpendicular to the Ising $z$ axis. 
As described below, the effective transverse field $\Gamma$ 
does not correspond one-to-one to the applied physical field $B_x$.
Rather, $\Gamma$ in Eq.~(\ref{Htfim}) is a function of the real
physical field $B_x$. 
In fact, $J_{ij}$ is also a function of $B_x$ ~\cite{Chakraborty,Tabei-MC,TabeiVernay}.
If the spin interactions, $J_{ij}$,  possess translational invariance,
the system displays for $\Gamma=0$ conventional long range magnetic order  below
some critical temperature, $T_{c}$.
In the simplest scenario, where $J_{ij}>0$, the ordered phase is ferromagnetic
and the order parameter is the average magnetization per spin,
$m_z=(1/N)\sum_{i}\langle S_{i}^{z}\rangle$, where $N$ is the number
of spins. Since $S_{i}^{x}$ and $S_{i}^{z}$ do not commute, nonzero
$\Gamma$ causes quantum tunneling between the spin-up, $\ket{\uparrow}$,
and spin-down, $\ket{\downarrow}$, eigenstates of $S_{i}^{z}$.
By increasing $\Gamma$, $T_{c}$ decreases until, ultimately, $T_{c}(\Gamma)$
vanishes at a \textit{quantum critical point} where $\Gamma=\Gamma_{c}$.
On the $T=0$ temperature axis, the system is in a long-range ordered
phase for $\Gamma<\Gamma_{c}$ while it is in a quantum paramagnetic
phase for $\Gamma>\Gamma_{c}$. The phase transition between the paramagnetic
and long range ordered phase at $\Gamma_{c}$ constitutes the quantum
phase transition~\cite{Pfeuty,Sen}.

One can also consider generalizations of the $H_{{\rm TFIM}}$ where the
$J_{ij}$ are quenched (frozen) random interactions. Of particular
interest is the situation where there are as many ferromagnetic $J_{ij}>0$
and antiferromagnetic $J_{ij}<0$ couplings. This causes a high level of
random frustration and the system, provided it is three dimensional, freezes
into a spin glass state via a true thermodynamic phase transition
at a spin glass critical temperature $T_{g}$~\cite{Binder,Mydosh}.
Here as well, one can investigate how the spin glass transition is
affected by a transverse field $\Gamma$. As in the previous example,
$T_{g}(\Gamma)$ decreases as $\Gamma$ is increased from zero until,
at $\Gamma=\Gamma_{c}$, a quantum phase transition between 
a quantum paramagnet and a spin glass phase ensues. Extensive numerical
studies have found the quantum phase transition between a quantum
paramagnet and a spin glass phase~\cite{Rieger-PRL,Rieger-PRB,Guo} 
to be quite interesting due to the
occurrence of Griffiths-McCoy singularities (GMS)~\cite{Griffiths,McCoY}.
 These GMS arise
from rare spatial regions of disorder which may, for example, resemble
the otherwise non-random (disorder-free) version of the system at
stake. As a result, GMS can lead to singularities in various thermodynamic
quantities away from the quantum critical point.

\subsection{LiHo$_x$Y$_{1-x}$F$_4$}

On the experimental side, most studies aimed at exploring the phenomena
associated with the TFIM have focused on the insulating LiHo$_{x}$Y$_{1-x}$F$_{4}$
(LHYF) Ising magnetic 
material~\cite{Reich,Wu-sg,Wu-chi3,Wu-thesis,Bitko,Brooke,Brooke-thesis,Ghosh-Science,Ghosh-Nature,Ronnow,Ronnow2007,SilevitchNature,SilevitchPRL,Jonsson,SilevitchSG,Jonsson-new,recent}.
In this system, the Ho$^{3+}$
Ising spin direction is parallel to the $c$ axis of the body-centered
tetragonal structure of LHYF. 
The random disorder is introduced by diluting the magnetic Ho$^{3+}$ ions by
non-magnetic Y$^{3+}$.
Crystal field effects lift the degeneracy of the $^5$I$_8$ electronic manifold,
giving an Ising ground doublet $\vert\Phi_0^\pm \rangle$
and a first excited $\vert\Phi_e \rangle$ singlet at approximately 11 K above the 
ground doublet~\cite{Ronnow2007}. 
The other 14 crystal field states lie at much higher energies~\cite{Ronnow2007}.
Quantum spin flip fluctuations are introduced by the application of a 
magnetic field, $B_x$, perpendicular to the Ising $c$ axis.
$B_x$ admixes  $\vert\Phi_e \rangle$  with 
 $\vert\Phi_0^\pm \rangle$, splitting the latter and producing an effective TFIM
with $\Gamma(B_x)\propto B_x^2$ for small $B_x$~\cite{Chakraborty}.

The properties of pure LiHoF$_4$ in a transverse $B_x$ are now generally 
qualitatively well understood~\cite{Chakraborty}. 
Indeed, a recent quantum Monte Carlo study~\cite{Chakraborty} found general agreement between 
experiments and 
a microscopic model of LiHoF$_4$. However,
some quantitative discrepancies between 
Monte Carlo and experimental data, even near the classical paramagnetic to 
ferromagnetic transition where $B_x/T_c$ is small, 
do exist ~\cite{Chakraborty,Tabei-MC}.
One noteworthy effect at play in LHYF
 at low temperatures is the significant enhancement
of the zero temperature critical $B_x$, $B_x^c$, 
caused by the strong hyperfine nuclear interactions in Ho$^{3+}$-based
materials~\cite{Bitko,Ronnow,Chakraborty,Schechter-PRL1}.

LiHo$_x$Y$_{1-x}$F$_4$ in a transverse $B_x$ and $x<1$ 
has long been known to display  paradoxical behaviors, both in the 
ferromagnetic (FM) ($0.25 < x <1.0$) and spin glass (SG) ($x<0.25$) regimes.
In the FM regime, a mean-field behavior $T_c(x) \propto x$ for 
the PM to FM transition is observed when $B_x=0$ ~\cite{Brooke}.
However, in nonzero $B_x$, the rate at which
$T_c(B_{x})$ is reduced by $B_{x}>0$ increases 
faster than mean-field theory predicts as $x$ is reduced~\cite{Brooke-thesis,SilevitchNature}.
In the high Ho$^{3+}$ (SG) dilution regime (e.g. LiHo$_{0.167}$Y$_{0.833}$F$_4$), LHYF
has long been~\cite{Wu-sg,Wu-chi3,Jonsson,recent}
argued to display a conventional SG transition for $B_x=0$ signalled by 
a nonlinear magnetic susceptibility, $\chi_3$, diverging at
$T_g$ as 
\mbox{$\chi_3(T)\propto (T-T_g)^{-\gamma}$ $\,$ \cite{Mydosh}}.
However, $\chi_3(T)$ becomes less singular as $B_{x}$ is increased from ${B_{x}=0}$,
suggesting that no quantum phase transition between a PM and a SG
state exists as 
\mbox{$T\rightarrow 0$ $\,$ \cite{Wu-chi3,Wu-thesis}}.
Recent theoretical studies~\cite{Schechter-PRL2,Tabei-PRL,TabeiVernay} 
suggest that for dipole-coupled Ho$^{3+}$ in a diluted sample,
nonzero $B_{x}$ generates  longitudinal 
(along the Ising $\hat z$ direction) random fields 
that (i) lead to a faster decrease of $T_c(B_x)$ in the FM 
regime~\cite{Brooke-thesis,Tabei-PRL,SilevitchNature} and (ii) destroy the 
PM to SG transition for samples that otherwise
show a SG transition when 
$B_x=0$ 
\cite{Wu-chi3,Wu-thesis,Jonsson,Jonsson-new,recent,Tabei-PRL,TabeiVernay,Schechter-PRL2},
or, at least, lead to a disappearance of the $\chi_3$ 
divergence as $B_x$ is increased from zero~\cite{Wu-chi3,Wu-thesis,Tabei-PRL}.

Perhaps most interesting among the phenomena exhibited by LHYF is the
one referred to as {\it antiglass} and which has been predominantly 
investigated in
 LiHo$_{0.045}$Y$_{0.955}$F$_4$ ~\cite{Reich,Ghosh-Nature,Ghosh-Science,Quilliam,Jonsson}.
The reason for this name comes from
AC susceptibility data on LiHo$_{0.045}$Y$_{0.955}$F$_{4}$ which show
that the distribution of relaxation times {\it narrows} upon cooling below
300 mK ~\cite{Reich,Ghosh-Science,Ghosh-Nature}.
This behavior is quite different from that observed in conventional
spin glasses where the distribution of relaxation times broadens upon
approaching a spin glass transition at $T_{g}>0$ ~\cite{Binder,Mydosh}. 
The antiglass behavior has been interpreted as evidence that
the spin glass transition in  LiHo$_{x}$Y$_{1-x}$F$_4$ disappears
at some nonzero $x_c>0$. 
Results from more recent experimental studies on LiHo$_{0.165}$Y$_{0.835}$F$_4$ ($x=$ 16.5\%) 
and LiHo$_{0.045}$Y$_{0.955}$F$_{4}$ ($x=$ 4.5\%) suggest an absence of a genuine spin glass 
transition, even for a concentration of Ho as large as 16.5\% ~\cite{Jonsson,recent}.
In particular, it is in stark contrast with theoretical arguments~\cite{Stephen}
which predict that, because of the long-ranged $1/r^3$ nature of dipolar
interactions,  classical dipolar Ising spin glasses should have 
$T_g(x)>0$ for all $x>0$.
However, even more recent work asserts that there is indeed a thermodynamic
SG transition for $x=16.5\%$ ~\cite{SilevitchSG}, but
that the behavior found in 
LiHo$_{0.045}$Y$_{0.955}$F$_4$ is truly unconventional~\cite{SilevitchSG}.

Two very different scenarios for the failure of LiHo$_{0.045}$Y$_{0.955}$F$_{4}$
to show a spin glass transition have been put forward~\cite{Ghosh-Nature,Snyder,Biltmo1,Biltmo2}.
Firstly, it has been suggested that the (small) off-diagonal part of the
dipolar interactions lead to virtual crystal field excitations that
admix $\vert \Phi_0^\pm\rangle$ with $\vert \Phi_{\rm e}\rangle$
and give rise to non-magnetic singlets for spatially
close pairs of Ho$^{3+}$ ions.
The formation of these singlets would
thwart the development of a spin glass state.
This mechanism is analogous to the one leading to the formation
of the random singlet state in dilute antiferromagnetically coupled
$S=1/2$ Heisenberg spins~\cite{Bhatt}.
However, a recent study~\cite{Chin} shows that the energy scale for this singlet
formation is very low ($\sim 10^0$ mK) and that the random singlet
mechanism~\cite{Ghosh-Nature}  may not be very effective at 
destroying the spin glass state in LiHo$_{0.045}$Y$_{0.955}$F$_4$ ~\cite{Ghosh-Nature}.
Hence the proposed formation of an entangled state 
in LiHo$_{0.045}$Y$_{0.955}$F$_4$ may, if it really exist, 
perhaps proceed via 
a more complex scheme than that proposed in Ref.~[\onlinecite{Ghosh-Nature}].
Also, the low-temperature features observed in the specific 
heat  in  Ref.~[\onlinecite{Ghosh-Nature}] have not been observed in a more
recent study~\cite{Quilliam}.
Secondly, and from a completely different perspective,  numerical simulations
of {\it classical} Ising dipoles found that the spin glass transition
temperature, $T_{g}$ appears to vanish for a concentration of dipoles below
approximately 20\% of the sites occupied~\cite{Snyder,Biltmo1,Biltmo2}.
However, even more recent Monte Carlo simulations find that this conclusion
may not be that firmly established~\cite{Tam}.

As another possible and yet unexplored 
scenario, we note here that since Ho$^{3+}$ is an even electron
system (i.e. a non-Kramers ion), the Kramers' theorem is inoperative
and the ground state doublet can be split by random (electrostatic)
crystal field effects that compete with the collective spin glass behavior. For
example, random strains, which may come from the substitution of 
Ho$^{3+}$$\rightarrow$Y$^{3+}$,
break the local tetragonal symmetry and introduces (random) crystal
fields operators (e.g. $O_{2}^{\pm2}$) which have nonzero
matrix elements between the two states
$\vert\Phi_0^+ \rangle$ with $\vert \Phi_0^- \rangle$ of the ground doublet, 
splitting it, 
and possibly destroying the spin glass phase at low Ho$^{3+}$ concentration. 
Indeed, such random transverse fields have been identified in 
samples with very dilute Ho$^{3+}$ in a LiYF$_4$ matrix~\cite{Sharukov,Bertaina}.
Also, very weak random strains, hence effective
random transverse fields, arise from the different (random) anharmonic zero point motion of 
$^6$Li and $^7$Li in Ho:LiYF$_4$ samples with natural abundance
of  $^6$Li and $^7$Li ~\cite{Agladze}. 
Finally, there may be intrinsic strains in the crystalline samples
that do not arise from the Ho$^{3+}$/Y$^{3+}$ or 
$^6$Li/$^7$Li admisture~\cite{Sharukov}.
However, using available estimates~\cite{Sharukov,Bertaina,Agladze},
calculations suggests that strain-induced random fields 
at play in LiHo$_{0.045}$Y$_{0.955}$F$_{4}$ 
may be too small [$<O(10^1~{\rm mK})$] to cause
the destruction of the spin glass phase in this system~\cite{Fortin}.
Nevertheless, the point remains that, 
in principle, the non-Kramers nature of Ho$^{3+}$  
does offer a route for the destruction
of the spin glass phase in LiHo$_x$Y$_{1-x}$F$_4$ outside strictly pairwise, 
quantum~\cite{Ghosh-Nature,Chin} or classical~\cite{Snyder,Biltmo1,Biltmo2},
magnetic interaction mechanisms. 
At this stage, this is clearly 
a matter that needs to be investigated experimentally further.
One notes that, because of Kramers' theorem, 
the destruction of a SG phase via strain-induced effective random transverse fields
would not occur for an odd-electron (Kramers) ion such as Dy$^{3+}$ or Er$^{3+}$.
In that context, one might think that a comparison of the behavior
of LiDy$_{x}$Y$_{1-x}$F$_{4}$ or LiEr$_{x}$Y$_{1-x}$F$_{4}$ 
with that of LiHo$_{x}$Y$_{1-x}$F$_{4}$
would be interesting. Unfortunately, while LiHo$_{x}$Y$_{1-x}$F$_{4}$
is an Ising system, the Er$^{3+}$ and Dy$^{3+}$ moments in 
LiEr$_{x}$Y$_{1-x}$F$_{4}$ and LiDy$_{x}$Y$_{1-x}$F$_{4}$
are XY-like~\cite{Magarino,Kazei}.
Hence, one cannot  compare the 
LiEr$_{x}$Y$_{1-x}$F$_{4}$ and LiDy$_{x}$Y$_{1-x}$F$_{4}$ XY compounds
with the 
LiHo$_{x}$Y$_{1-x}$F$_{4}$ Ising material on the same footing.

From the above discussion, it is clear that there are a number of fundamental
questions raised by experimental studies of LiHo$_x$Y$_{1-x}$F$_4$, 
both in zero and nonzero transverse field $B_x$,
that warrant systematic experimental investigations in other 
similar diamagnetically-diluted dipolar Ising-like magnetic materials. 
Specific questions are:

\begin{enumerate}
\item  How does the quantum criticality of a transverse field Ising material 
	  with much smaller hyperfine interactions than Ho$^{3+}$ in
	 LiHo$_x$Y$_{1-x}$F$_4$ manifest itself~\cite{Bitko,Ronnow,Schechter-PRL1} ?
\item Is the theoretical proposal of transverse-induced random longitudinal fields
in diluted dipolar Ising materials~\cite{Tabei-PRL,Schechter-PRL2,TabeiVernay}
valid and can it be explored and confirmed in other materials other than
LiHo$_x$Y$_{1-x}$F$_4$ \cite{SilevitchNature}?
In particular, are the phenomena observed in Ref.~[\onlinecite{SilevitchNature}] and
ascribed to Griffiths singularities observed in other
disordered dipolar Ising systems subject to a transverse field?
\item Does the antiglass phenomenon~\cite{Reich,Ghosh-Science,Ghosh-Nature} 
occur in other diluted dipolar Ising materials? If yes, which, 
if any, of the aforementioned theoretical proposals 
for the destruction of the spin glass state
at small spin concentration is correct?
\end{enumerate}

\subsection{RE(OH)$_3$ materials}

As mentioned above, these questions cannot be investigated with the
LiEr$_{x}$Y$_{1-x}$F$_{4}$ and LiDy$_{x}$Y$_{1-x}$F$_{4}$ materials isotructural
to the LiHo$_x$Y$_{1-x}$F$_4$ Ising compound since they are XY-like systems.
However, we note in passing that it would nevertheless be interesting to explore
the topic of induced random fields~\cite{Schechter-PRL2,Tabei-PRL,TabeiVernay} 
and the possible existence of an XY dipolar spin glass
and/or antiglass state in LiEr$_{x}$Y$_{1-x}$F$_{4}$ and LiDy$_{x}$Y$_{1-x}$F$_{4}$.
The LiTb$_{x}$Y$_{1-x}$F$_{4}$ material is of limited  use in such investigations 
since the single ion ground state of Tb$^{3+}$ in this compound consists of two
separated singlets~\cite{LiTbF4-singlet}, 
and local moment magnetism on the Tb$^{3+}$ site disappears at low Tb concentration~\cite{Aeppli-Nato}.
In this paper, we propose that the RE(OH)$_3$ (RE=Ho, Dy) compounds may offer themselves
as an attractive class of materials to study the above questions.
Similarly to the LiHoF$_4$, the RE(OH)$_3$ materials possess the following interesting properties:

\begin{enumerate}

\item They are insulating rare earth materials.
\item Their main spin-spin couplings are magnetostatic dipole-dipole interactions.
\item The RE(OH)$_3$ materials are stable at room temperature.
\item Both pure RE(OH)$_3$ and LiHoF$_4$ are colinear (Ising-like) dipolar 
ferromagnets with the Ising direction along the $c$ axis 
of a hexagonal unit cell (RE(OH)$_3$) or body-centered tetragonal unit cell (LiHoF$_4$).
In both cases there are two magnetically equivalent ions per unit cell.
\item In  RE(OH)$_3$, the Kramers (Dy$^{3+}$) and non-Kramers (Ho$^{3+}$) variants possess a common
crystalline structure and both have similar bulk magnetic properties in zero transverse 
magnetic field $B_x$.
\item The critical temperature of the pure RE(OH)$_3$ compounds is relatively high, $\sim 3~{\rm K}$.
This would make possible the study of
Y$-$substituted Dy and Ho hydroxides down to quite low concentration
of rare earth while maintaining the relevant magnetic temperature scale above 
the lowest attainable temperature with a commercial dilution refrigerator.
\item Finally, and this is a key feature that motivated the present study,
the first excited crystal field state in the 
Ho(OH)$_3$ and Dy(OH)$_3$ compounds is low-lying, hence allowing a possible
transverse-field induced admixing and, possibly, 
a transverse field Ising model description~\cite{spin-ice}

\end{enumerate}

To the best of our knowledge, it appears that the RE(OH)$_3$ materials
have so far not been investigated as potential realization of the TFIM.
The purpose of this paper is to explore (i) the possible description of these
materials as a TFIM, (ii) obtain an estimate
of what the zero temperature critical transverse field $B_x^c$ 
may be and, (iii) assess if any new interesting phenomenology may occur, even in 
the pure compounds, in nonzero transverse field $B_x$.

We note, however, that there are so far no very large
single crystals of RE(OH)$_{3}$ available~\cite{Catanese-Tb}. 
For example, their length typically varies between 3 mm and 17 mm and their diameter
between 0.2 and 0.6 mm.
The lack of large single crystals would make difficult neutron scattering experiments.
However, possibly motivated by this work and by a first generation of bulk
measurements (e.g. susceptibility, specific heat), experimentalists
and solid state chemists 
may be able to conceive ways to grow larger single crystals of RE(OH)$_3$. 
Also, in light of the fact that most experiments on LiHo$_{x}$Y$_{1-x}$F$_{4}$
that have revealed exotic behavior are bulk 
measurements~\cite{Reich,Wu-sg,Wu-chi3,Brooke,Ghosh-Science,Ghosh-Nature,Quilliam},
we hope that at this time the lack of availability of large single
crystals of the RE(OH)$_{3}$ series is not a strong impediment against pursuing 
a first generation of bulk experiments on RE(OH)$_3$.

The rest of this paper is organized as follows.
In Section II, we review the main, single ion, magnetic properties of RE(OH)$_3$
(RE=Dy, Ho, Tb). In particular, we discuss the crystal field Hamiltonian of these
materials and the dependence of the low-lying crystal field levels 
on an applied transverse field $B_x$. 
We present in Section III a mean-field calculation to estimate the $B_x$ vs temperature, $T$, 
$T_c(B_x)$ phase diagram of these materials. 
In Section IV, we show that Ho(OH)$_3$ and Tb(OH)$_3$ 
can be described quantitatively well by a transverse field Ising model, while
Dy(OH)$_3$ cannot.
The Subsection A of Section V, uses a Ginzburg-Landau theory to explore the first order paramagnetic (PM) to
ferromagnetic (FM) transition that occurs in Dy(OH)$_3$ at low temperatures and strong
$B_x$. The following Subsection B discusses the effect of nearest-neighbor antiferromagnetic exchange interaction
and applied longitudinal (i.e. along the $\hat z$ axis) magnetic field, $B_z$, on the first
order transition in Dy(OH)$_3$.
A brief conclusion is presented in Section VI.
Appendix A discusses how the excited crystal field states in Dy(OH)$_3$ play an important
quantitative role on the determination of $T_c(B_x)$ in this material.

\section{RE(OH)$_{{\bf 3}}$: Material properties\label{Materials}}

\subsection{Crystal properties}

\label{Mat-prop}

\begin{figure}
\includegraphics[width=1.0\columnwidth,keepaspectratio]{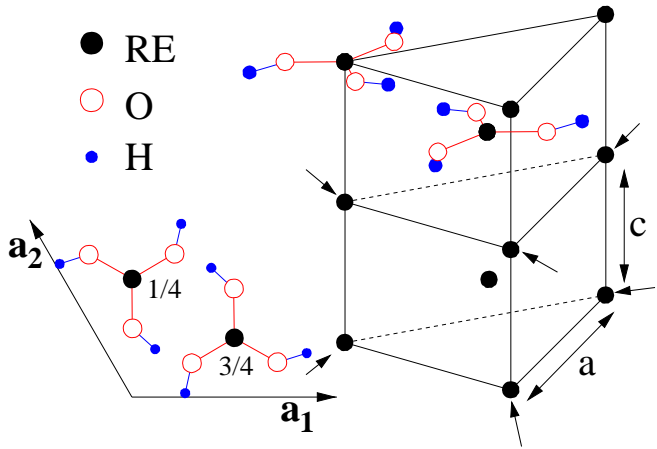}
\caption{The lattice structure of rare earth hydroxides and Y(OH)$_{3}$. 
The arrows indicate the 6 nearest neighbors of the central ion in the lower prism.
 The lower left corner shows a projection of the unit cell onto 
the plane given by lattice vectors  $\bm{a_1}$ and  $\bm{a_2}$. 
\label{fig:Lattice}}
\end{figure}

All the rare earth hydroxides form hexagonal crystals  that 
that 
are iso-structural
with Y(OH)$_{3}$. 
The lattice is described by translation vectors $\bm{a_1}=(0,0,0)$, 
$\bm{a_2}=(-a/2,a\sqrt{3}/2,0)$ and $\bm{a_3}=(0,0,c)$.
A unit cell consist of two Ho$^{3+}$ ions at coordinates $(1/3, 2/3, 1/4)$ and $(2/3, 1/3, 3/4)$ 
in the basis of lattice vectors $\bm{a_1}$, $\bm{a_2}$ and $\bm{a_3}$. 
The coordinates of three O$^{2-}$ and H$^{-}$ ions, 
relative to the position of Ho$^{3+}$, 
are \mbox{$\pm(x, y, 0)$}, \mbox{$\pm(-y, x-y, 0)$} and \mbox{$\pm(y-x, -x, 0)$}, 
where $\pm$ refers to the first and second Ho$^{3+}$ in the unit cell, respectively~\cite{Scott-thesis}.
The values of the parameters x and y are listed in Table \ref{tab:Lattice-xy}.
x=0.396, y=0.312 and for H$^{-}$: x=0.28, y=0.17\cite{Beall}. 
The lattice structure is depicted in Fig. \ref{fig:Lattice}.  
The lattice
constants for Tb(OH)$_{3}$, Dy(OH)$_{3}$, Ho(OH)$_{3}$ and Y(OH)$_{3}$,
from Beall \textit{et al}. \cite{Beall}, are collected
in Table \ref{tab:Latice-const}. Each magnetic ion is surrounded
by 9 oxygen atoms that create a crystalline field characterized by
the point group symmetry  $C_{3h}$\cite{Scott-thesis}. 

\begin{table}
\begin{tabular}{|c|c|c|c|c|}
\hline
&
O$^{2-}$, $x$&
O$^{2-}$, $y$&
H$^{-}$, $x$&
H$^{-}$, $y$\tabularnewline
\hline
\hline
Tb(OH)$_{3}$&
$0.3952(7)$&
$0.3120(6)$&
$0.276(1)$&
$0.142(1)$\tabularnewline
\hline
Dy(OH)$_{3}$&
$0.3947(6)$&
$0.3109(6)$&
$0.29(3)$&
$0.15(2)$\tabularnewline
\hline
Ho(OH)$_{3}$&
$0.3951(7)$&
$0.3112(7)$&
$0.30(3)$&
$0.17(3)$\tabularnewline
\hline
Y(OH)$_{3}$&
$0.3958(6)$&
$0.3116(6)$&
$0.28(1)$&
$0.17(1)$\tabularnewline
\hline
\end{tabular}

\caption{Position parameters of  O$^{2-}$ and H$^{-}$ ions in rare earth
hydroxides and Y(OH)$_{3}$ from Ref.~[\onlinecite{Beall}] and
Ref.~[\onlinecite{Landers}] (see text in Section \ref{Mat-prop}).
\label{tab:Lattice-xy}
}
\end{table}

\begin{table}
\begin{tabular}{|c|c|c|c|}
\hline 
&
$a$&
$c$&
$c/a$\tabularnewline
\hline
\hline 
Tb(OH)$_{3}$&
$6.315(4)$&
$3.603(2)$&
$0.570(5)$\tabularnewline
\hline 
Dy(OH)$_{3}$&
$6.286(3)$&
$3.577(1)$&
$0.569(0)$\tabularnewline
\hline 
Ho(OH)$_{3}$&
$6.266(2)$&
$3.553(1)$&
$0.567(0)$\tabularnewline
\hline 
Y(OH)$_{3}$&
$6.261(2)$&
$3.544(1)$&
$0.566(0)$\tabularnewline
\hline
\end{tabular}
\caption{
Lattice constants for rare earth hydroxides and 
Y(OH)$_{3}$ (from Ref.~[\onlinecite{Beall}]).\label{tab:Latice-const}}
\end{table}

\subsection{Single ion properties}

The electronic configuration of the magnetic ions Tb$^{3+}$, Dy$^{3+}$
and Ho$^{3+}$ is, respectively, $4f^{8}$, $4f^{9}$ and $4f^{10}$.
Magnetic properties of the rare earth ions can be described
by the states of the lowest energy multiplet: the spin-obit splitting
between the ground state $J$ manifold and the first excited states
is of order of few thousands K. The ground state manifolds can be
found from  Hund's rules and are $^{7}F_{6}$, $^{6}H_{15/2}$
and $^{5}I_{8}$ for Tb$^{3+}$, Dy$^{3+}$ and Ho$^{3+}$, respectively.
The Wigner-Eckart theorem gives the Land\'e $g$-factor equal to $\frac{3}{2}$,
$\frac{4}{3}$ and $\frac{5}{4}$ for Tb$^{3+}$, Dy$^{3+}$ and Ho$^{3+}$,
respectively.

In a crystalline environment, an ion is subject to the electric field
and covalency effects from the surrounding ions. 
This crystalline field effect partially lifts the
degeneracy of the ground state multiplet. The low energy
levels of Tb$^{3+}$ in Tb(OH)$_{3}$ are a pair of singlets,
that consist of the symmetric combination of $\ket{\pm6}$ states with
a small admixture of the $\ket{0}$ state and an antisymmetric combination
0.3 cm$^{-1}$ above~\cite{Scott-Tb}. The next excited state is well separated from
the lowest energy pair by 
an  energy of 118 cm$^{-1}$ ~\cite{Scott-Tb} (1 cm$^{-1} \approx 1.44$ K).
In the case of Dy$^{3+}$ in Dy(OH)$_{3}$, the spectrum consist of
8 Kramers doublets with the first excited state 7.8 cm$^{-1}$ above
the ground state \cite{Scott-Dy}. The low energy spectrum of Ho$^{3+}$
in Ho(OH)$_{3}$ is composed of  a ground state doublet and an excited singlet state
11.1 cm$^{-1}$ above \cite{Scott-thesis}.

Due to the strong shielding of the $4f$ electrons by the electrons
of the filled outer electronic shells, the exchange interactions for $4f$ electrons 
is weak and the crystal field can be considered
as a perturbation to the fixed $J$ manifold. Furthermore, because the
strong spin-obit interaction yields a large energy gap between the ground
state multiplet and the excited levels, we neglect all the excited
electronic multiplets in the calculation.

According to arguments provided by Stevens~\cite{Stevens}, we 
express the matrix elements of the crystal field Hamiltonian for the
ground state manifold in terms of operator equivalents. The details
of the method and conventions for expressing the crystal field Hamiltonian
can be found in the review by Hutchings~\cite{Hutchings}. 
On the basis of the Wigner-Eckart theorem,
one can write the crystal field Hamiltonian in the form
\begin{equation}
\mathcal{H}_{\rm cf}=\sum_{nm}\theta_{n}B_{n}^{m}O_{n}^{m},
\label{eq:genCFH}
\end{equation}
 where $O_{n}^{m}$ are Steven's 
``operator equivalents'',
$\theta_{n}$ are constants called Stevens multiplicative factors
and $B_{n}^{m}$ are crystal field parameters (CFP). 
The CFP are usually
determined by fitting experimental (spectroscopic) data. From  angular momentum algebra, we
know that in the case of f electrons, we need to consider only $n=0,2,4,6$
 in the sum (\ref{eq:genCFH}). 
The choice of $B_{n}^{m}$ coefficients in Hamiltonian (\ref{eq:genCFH}) that 
do not vanish and have nonzero corresponding matrix elements 
is dictated by the point symmetry group of the
crystalline environment. The Stevens operators, $O_{n}^{m}$, are conveniently
expressed in terms of vector components of angular momentum operator
${\bm J}$. In the case of the RE(OH)$_3$ materials,
considered herein, the point-symmetry group
is $C_{3h}$, and the crystal field Hamiltonian is of the form 
\begin{eqnarray}
{\cal H}_{\rm cf}({{\bm J}}_{i}) & = & 
\alpha_{J}B_{2}^{0}O_{2}^{0}({{\bm J}}_{i})+
\beta_{J}B_{4}^{0}O_{4}^{0}({{\bm J}}_{i})\nonumber \\
 &  & 
+\gamma_{J}B_{6}^{0}O_{6}^{0}({{\bm J}}_{i})+
\gamma_{J}B_{6}^{6}O_{6}^{6}({{\bm J}}_{i}).
\label{eq:Vc}
\end{eqnarray}
 The Stevens multiplicative factors $\alpha_{J}$, $\beta_{J}$ and
$\gamma_{J}$ ($\theta_{2}$, $\theta_{4}$ and $\theta_{6}$) are
collected in Table. \ref{tab:Stevens-multiplicative-factors}.

\begin{table*}
\begin{tabular}{|c|c|c|c|}
\hline 
\multicolumn{1}{|c|}{{\footnotesize Ion}}&
{\footnotesize $\alpha_{J}$}&
{\footnotesize $\beta_{J}$ }&
{\footnotesize $\gamma_{J}$}\tabularnewline
\hline
\hline 
{\footnotesize Tb$^{3+}$}&
{\footnotesize $-1/(3^{2}\cdot11)$}&
{\footnotesize $2/(3^{3}\cdot5\cdot11^{2})$}&
{\footnotesize $-1/(3^{4}\cdot7\cdot11^{2}\cdot13)$}\tabularnewline
\hline 
{\footnotesize Dy$^{3+}$}&
{\footnotesize $-2/(3^{2}\cdot5\cdot7)$}&
{\footnotesize $-2^{3}/(3^{3}\cdot5\cdot7\cdot11\cdot13)$}&
{\footnotesize $2^{2}/(3^{3}\cdot7\cdot11^{2}\cdot13^{2})$}\tabularnewline
\hline 
{\footnotesize Ho$^{3+}$}&
{\footnotesize $-1/(2\cdot3^{2}\cdot5^{2})$}&
{\footnotesize $-1/(2\cdot3\cdot5\cdot7\cdot11\cdot13)$}&
{\footnotesize $-5/(3^{3}\cdot7\cdot11^{2}\cdot13^{2})$}\tabularnewline
\hline
\end{tabular}

\caption{Stevens multiplicative factors\label{tab:Stevens-multiplicative-factors}}
\end{table*}

\begin{table*}
\begin{tabular}{|c|c|c|c|c|c|}
\hline 
{\small Ref}&
{\small Crystal}&
{\small $B_{2}^{0}$(cm$^{-1}$)}&
{\small $B_{4}^{0}$(cm$^{-1}$)}&
{\small $B_{6}^{0}$(cm$^{-1}$)}&
{\small $B_{6}^{6}$(cm$^{-1}$)}\tabularnewline
\hline
\hline 
{\small \cite{Scott-thesis}}&
{\small Tb(OH)$_{3}${*}}&
{\small $207.9\pm2.8$}&
{\small $-69.0\pm1.6$}&
{\small $-45.3\pm1.1$}&
{\small $585\pm10$}\tabularnewline
\hline 
{\small \cite{Scott-thesis}}&
{\small Tb:Y(OH)$_{3}$}&
{\small $189.1\pm2.6$}&
{\small $-69.1\pm1.5$}&
{\small $-45.7\pm1.0$}&
{\small $606\pm9$}\tabularnewline
\hline 
{\small \cite{Scott-thesis}}&
{\small Dy(OH)$_{3}${*}}&
{\small $209.4\pm3.4$}&
{\small $-75.5\pm3.5$}&
{\small $-40.1\pm1.9$}&
{\small $541.8\pm26.5$}\tabularnewline
\hline 
{\small \cite{Karmakar-Ho}}&
{\small Ho(OH)$_{3}$}&
{\small $200\pm2.0$}&
{\small $-57\pm0.5$}&
{\small $-40\pm0.5$}&
{\small $400\pm5$}\tabularnewline
\hline 
{\small \cite{Scott-thesis}}&
{\small Ho:Y(OH)$_{3}${*}}&
{\small $246.0\pm3.4$}&
{\small $-56.7\pm1.2$}&
{\small $-39.8\pm0.3$}&
{\small $543.6\pm3.3$}\tabularnewline
\hline 
\cite{Karmakar-Dy}&
{\small Dy(OH)$_{3}$}&
{\small $215.9$}&
{\small $-72.0$}&
{\small $-42.0$}&
{\small $515.3$}\tabularnewline
\hline
\end{tabular}

\caption{\label{tab:CFP}Crystal field parameters. 
Some of the calculation we performed with one set of crystal field parameters only. 
Crystal field parameters arbitrary chosen~\cite{Arbitrary}
 to be used in these calculations are marked with *.}
\end{table*}

For the sake of conciseness, and to illustrate the procedure,
most of our numerical results below are presented for 
one set of CFP only. 
The qualitative picture that emerges from our calculatiosn
does not depend on the
specific choice of CFP parameters. Only quantitative differences are found using
the different sets of CFP.
Ultimately, a further experimental determination of accurate
$B_n^m$ values would need to be carried out in order to obtain more precise
mean-field estimates as well as to perform 
quantum 
Monte Carlo simulations of
the Re(OH)$_3$ systems.
According to our {\it arbitrary} choice~\cite{Arbitrary}, if not stated otherwise, 
we use in the calculations the CFP provided by 
Scott \textit{et al.}~\cite{Scott-thesis,Scott-Tb-2,Scott-Dy,Scott-Tb}.
For Ho(OH)$_{3}$ and Dy(OH)$_{3}$ different values of CFP were proposed
by Karmakar {\it et al.}~\cite{Karmakar-Ho,Karmakar-Dy}. 
As one can see in Fig. \ref{fig:Phase-diagrams}, for Ho(OH)$_{3}$, 
the latter set of CFP yields a somewhat 
higher mean-field critical temperature and quite a bit 
higher critical value of the transverse magnetic
field $B_{x}^{c}=$7.35 T, compared with $B_{x}^{c}=$4.35 T obtained
using  Scott {\it et al.}'s CFP~\cite{Scott-thesis}(see Fig. \ref{fig:Phase-diagrams}). 
Similarly, Karmakar  {\it et al.}'s CFP ~\cite{Karmakar-Dy}
 for Dy(OH)$_{3}$ give a 
much higher critical field of $B_{x}^{c}=$9.12 T, compared with $B_{x}^{c}=$ 5.03 T
when Scott 's {\it et al.}'s CFP~\cite{Scott-thesis,Scott-Tb-2,Scott-Dy} are used.
From the two sets of CFP for Tb(OH)$_{3}$ we choose the
one obtained from measurements on pure Tb(OH)$_{3}$~\cite{Scott-thesis}.
Using the CFP obtained for 
the system with a dilute concentration of Tb in a Y(OH)$_3$ matrix,
Tb:Y(OH)$_{3}$~\cite{Scott-thesis}, makes only a small change to the
value of critical transverse field; we obtained $B_{x}^{c}=$50.0 T
and $B_{x}^{c}=$54.8 T calculated using Tb:Y(OH)$_{3}$ and Tb(OH)$_{3}$
CFP, respectively (see  Fig. \ref{fig:Phase-diagrams}). 
Available values of the CFP are given in Table \ref{tab:CFP}.

\begin{table}

 \begin{tabular}{lc}
\hline 
\multicolumn{1}{c}{Eigenstate}&
Energy $[cm^{-1}]$ \tabularnewline
\hline
\hline 
Dy(OH)$_{3}$&
\tabularnewline
\hline 
$0.92\ket{\pm15/2}-0.15\ket{\pm3/2}+0.37\ket{\mp9/2}$&
\tabularnewline
$0.40\ket{\pm15/2}+0.20\ket{\pm3/2}-0.90\ket{\mp9/2}$&
9.6\tabularnewline
\hline
\hline 
Dy(OH)$_{3}$ Karmakar {\it et al.}, Ref.~[\onlinecite{Karmakar-Ho,Karmakar-Dy}]&

\tabularnewline
\hline 
$0.98\ket{\pm15/2}-0.09\ket{\pm3/2}+0.15\ket{\mp9/2}$&
\tabularnewline
$0.17\ket{\pm15/2}+0.22\ket{\pm3/2}-0.96\ket{\mp9/2}$&
19.3\tabularnewline
\hline
\hline 
Ho:Y(OH)$_{3}$&
\tabularnewline
\hline 
$0.94\ket{\pm7}+0.31\ket{\pm1}+0.15\ket{\mp5}$ &
\tabularnewline
$0.59\ket{6}+0.55\ket{0}+0.59\ket{-6}$&
12.7\tabularnewline
\hline
\hline 
Ho(OH)$_{3}$ Karmakar {\it et al.}, Ref.~[\onlinecite{Karmakar-Ho,Karmakar-Dy}]&
\tabularnewline
\hline 
$0.97\ket{\pm7}+0.24\ket{\pm1}+0.09\ket{\mp5}$&
\tabularnewline
$0.60\ket{6}+0.52\ket{0}+0.60\ket{-6}$&
23.6\tabularnewline
\hline
\hline 
Tb(OH)$_{3}$&
\tabularnewline
\hline 
$0.71\ket{6}+0.05\ket{0}+0.71\ket{-6}$&
\tabularnewline
$0.71\ket{6}-0.71\ket{-6}$&
0.49\tabularnewline
$0.99\ket{\pm5}+0.13\ket{\mp1}$&
122.06\tabularnewline
\hline
\hline 
Tb:Y(OH)$_{3}$&
\tabularnewline
\hline 
$0.71\ket{6}+0.06\ket{0}+0.71\ket{-6}$&
\tabularnewline
$0.71\ket{6}-0.71\ket{-6}$&
0.58\tabularnewline
$0.99\ket{\pm5}+0.16\ket{\mp1}$&
115.33\tabularnewline
\hline
\end{tabular}

\caption{\label{tab:Eigenstates}Eigenstates and energy levels calculated
with the crystal field parameters collected in Table \ref{tab:CFP}.
}
\end{table}

We show in Table \ref{tab:Eigenstates} the lowest eigenstates and
eigenvalues of the crystal field Hamiltonian (\ref{eq:Vc}). The calculated
energies are not in full agreement with the experimentally
determined values because
the CFP were fitted using all the observed optical transitions,
including transitions between diffent ${\bm J}$ manifolds~\cite{Scott-thesis}. 
Furthermore, the fitting procedure used
by Scott~\cite{Scott-thesis} includes perturbative admixing between manifolds
with the admixing
incorporated into effective Stevens multiplicative
factors $\alpha_{J}$, $\beta_{J}$ and $\gamma_{J}$ that slightly 
differ from those given in 
Table  ~\ref{tab:Stevens-multiplicative-factors}.

Given the uncertainty in the CFP, which ultimately lead to an uncertainty
of approximately $\sim 40\%$ on $B_x^c$ in Ho(OH)$_3$ and Dy(OH)$_3$, as
well as the nature of the mean-field calculations that we use and
which neglects thermal and quantum mechanical fluctuations, as well as for simplicity sake, 
we ignore here the effect of hyperfine coupling of the electronic
and nuclear magnetic moments. 
However, as shown for LiHo$_x$Y$_{1-x}$F$_4$, the important 
role of hyperfine interactions 
for  Ho$^{3+}$ on the precise
determination of $B_x^c$ must eventually 
be considered~\cite{Bitko,Chakraborty,Ronnow,Schechter-PRL1}.
At this time, one must await  results from further experiments 
and a precise set of CFP for ${\cal H}_c$ 
in order to go beyond the mean-field calculations presented below or to
pursue quantum Monte Carlo calculations as done in Refs.~[\onlinecite{Chakraborty,Tabei-MC}].
As suggested in Ref.~[\onlinecite{Chakraborty}], the accuracy of
any future calculations (mean-field or quantum Monte Carlo) could 
be improved by the use of 
directly measured 
accurate values of the transverse field splitting 
of the ground state doublet instead of the less certain values calculated from CFP.

Since our main goal in this exploratory work is to estimate the critical transverse
field, $B_x^c$, for the family of RE(OH)$_3$ compounds and to explore the possible validity
of a transverse field Ising model description of these materials, we 
henceforth restrict ourselves to the
${\cal H}_{\rm cf}$ in Eq.~(\ref{eq:genCFH}) with the 
CFP ($B_n^m$ parameter values) given in Table ~\ref{tab:CFP}.
These calculations could be revisited and quantum
 Monte Carlo simulations~\cite{Chakraborty,Tabei-MC}
 performed once experimental results reporting on the effect of $B_x$ on 
Dy(OH)$_3$ and Ho(OH)$_3$ become available.

\subsection{Single ion transverse field spectrum}

\begin{figure}[ht]
\includegraphics[width=1.0\columnwidth,keepaspectratio]{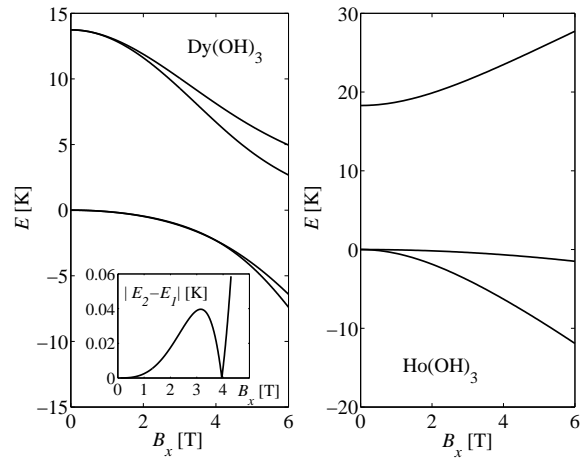}
\caption{Energy levels as a function of applied transverse field for Dy(OH)$_{3}$
and Ho(OH)$_{3}$ . The inset shows the separation of the lowest energy
levels in Dy(OH)$_{3}$.\label{fig:EnergyLevels}}
\end{figure}

A magnetic field, $B_{x}$, applied in the direction transverse to the
easy axis splits the degeneracy of the ground state doublet in the
case of Ho(OH)$_{3}$ and Dy(OH)$_{3}$, or increase the separation
of the ground levels in the case of 
the already weakly separated singlets in Tb(OH)$_{3}$. 
By diagonalizing the single-ion
Hamiltonian, ${\cal H}_0$,
which consist of the crystal field and Zeeman term, 
\begin{equation}
\mathcal{H}_{0}={\cal H}_{\rm cf}({\bm J}_{i})-g\mu_{\rm B}B_{x}J_{x}  ,
\label{eq:HZeeman}
\end{equation}
 we obtain the transverse field dependence of the single ion energy levels, plotted
in Fig. \ref{fig:EnergyLevels}. In the case of Dy(OH)$_{3}$, the
two lowest energy levels splitting is too small to be clearly visible
in the main panel of  Fig. \ref{fig:EnergyLevels}.
Hence, we show the energy separation between the two lowest levels in
the inset of Fig. \ref{fig:EnergyLevels} for Dy(OH)$_3$.
Furthermore, the separation vanishes at $B_{x}=3.92$ T, 
indicating that the two lowest states for this specific value
of the transverse field, $B_{x}$, are degenerate.

To calculate the transverse field dependence of the lowest energy levels 
up to the critical transverse field where dipolar ferromagnetism is
destroyed,
we do not have to include all the crystal field states since
the $B_x$-induced admixing 
among the states decreases with increasing energy separation. 
In the case of Ho(OH)$_{3}$ we can reproduce the field
dependence, $B_x$,  of the lowest energy levels, $E$ in
 Fig. \ref{fig:EnergyLevels} using only the four lowest levels. 
However, in order to achieve a similar level of agreement for
 Dy(OH)$_{3}$, we have to retain 
the ground doublet and 
several of the lowest excited doublets.

\section{Numerical solution\label{Numerics}}

\begin{figure}[ht]
\includegraphics[width=1.0\columnwidth,keepaspectratio]{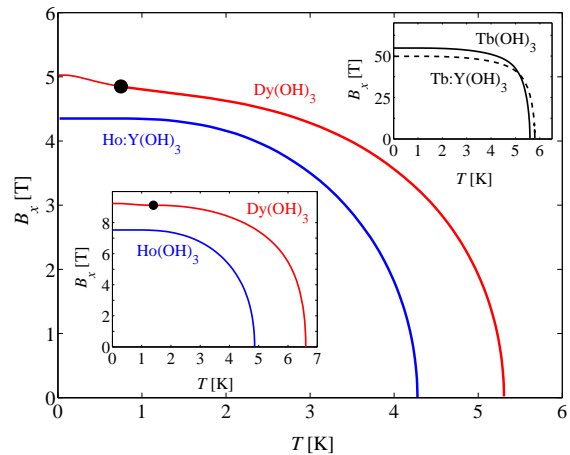}
\caption{The main panel shows the phase diagrams for Dy(OH)$_{3}$ and Ho:Y(OH)$_{3}$ (crystal field parameters of Scott {\it et al.}~\cite{Scott-thesis,Scott-Dy}). 
The dot indicates the location of the tricritical point for Dy(OH)$_3$. 
The transition is first order for temperatures below the temperature location of the
tricritical point.
The upper inset shows the phase diagram for Tb(OH)$_{3}$ (crystal field parameters of Scott {\it et al.}~\cite{Scott-thesis,Scott-Tb,Scott-Tb-2}).
The lower inset shows the phase diagram for Ho(OH)$_{3}$
and Dy(OH)$_{3}$ calculated with the crystal field parameters of Karmakar {\it et al.}~\cite{Karmakar-Ho,Karmakar-Dy}.
\label{fig:Phase-diagrams}}
\end{figure}

The collective  magnetic properties of the considered rare earth hydroxides
are mainly controlled by a long range dipolar interaction between 
the magnetic moments carried by the rare earth ions. 
The dipolar interaction is complemented by a short range exchange interaction. 
Adding the interaction terms to the single ion Hamiltonian (\ref{eq:HZeeman})
gives a full Hamiltonian, ${\cal H}$, of the form 
\begin{eqnarray}
\mathcal{H} & = & \sum_{i}{\cal H}_{\rm cf}({\bm J}_{i})-
g\mu_{\rm B}\sum_{i}B_{x}J_{i,x}\nonumber \\
 & + & \frac{1}{2}\sum_{ij}\sum_{\mu\nu}\mathcal{L}_{ij}^{\mu\nu}J_{i,\mu}J_{j,\nu}+
\frac{1}{2}
\mathcal{J_{{\rm ex}}}\sum_{i,nn}{\bm J}_{i} \cdot {\bm J}_{nn}.
\label{eq:Hfull}
\end{eqnarray}
$\mathcal{L}_{ij}^{\mu\nu}$ 
are the anisotropic dipole-dipole interaction constants
of the form 
$\mathcal{L}_{ij}^{\mu\nu}=\frac{\mu_{0}(g\mu_{\rm B})^{2}}{4\pi a^{3}}L_{ij}^{\mu\nu}$,
where $\mu,\nu=x,y,z$; $a$ is a lattice constant 
(see Table~\ref{tab:Latice-const})
and $\mu_{0}$ is the permeability of vacuum. $L_{ij}^{\mu\nu}$ are dimensionless
dipolar interaction coefficients,
\begin{equation}
L_{ij}^{\mu\nu}=\frac{\delta^{\mu\nu}
\left| {\bm r}_{ij}\right|^{2}-3\left( {\bm r}_{ij}\right)^{\mu}
\left( {\bm r}_{ij}\right)^{\nu}}{\left| {\bm r}_{ij}\right|^{5}},
\label{eq:Lij}
\end{equation} 
where ${\bm r}_{ij}={\bm r}_{j}-{\bm r}_{i}$, with ${\bm r}_{i}$ 
the lattice position of magnetic moment ${\bm J_i}$, 
expressed in units of the lattice constant, $a$.
$\mathcal{J_{{\rm ex}}}$ is the 
antiferromagnetic ($\mathcal{J_{{\rm ex}}}>0$) exchange interaction constant, 
which can be recast 
as $\mathcal{J_{{\rm ex}}}=\frac{\mu_{0}(g\mu_{\rm  B})^{2}}{4\pi a^{3}}J_{{\rm ex}}$,
where $J_{{\rm ex}}$ is now a dimensionless exchange constant
that, when multiplied by the nearest neighbor
coordination number, $z$=6, 
can be use to compare
the relative strength of exchange vs the magnetic 
dipolar lattice sum (energies) 
collected in Table
\ref{tab:Ls}. The label $nn$ in Eq.~(\ref{eq:Hfull})
denotes the  nearest neighbor sites of  site $i$.

The exchange interaction is expected to be 
of somewhat lower strength than the dipolar coupling~\cite{Catanese-DyHo, Catanese-Tb}. 
We therefore neglect it in most of the calculations, 
but we discuss its effect on the calculated $B_x$ 
vs $T$ phase diagram at the end of this section
as well as explore its influence on the occurrence 
of a first order phase transition in Dy(OH)$_{3}$ in Section \ref{sec:1stOrdBzJex}.
Denoting
$L_{\mu\nu}=\sum_{j}L_{ij}^{\mu\nu}$ and 
$\mathcal{L}_{\mu\nu}=\frac{\mu_{0}(g\mu_{\rm B})^{2}}{4\pi a^{3}}L_{\mu\nu}$,
we write a mean-field Hamiltonian in the form
\begin{eqnarray}
\mathcal{H}_{\rm MF} & = & {\cal H}_{c}({\bm J})-g \mu_{\rm B}B_{x}J_{x} \nonumber  \\
 & + & \sum_{\mu=x,y,z} 
\left(  \mathcal{L}_{\mu\mu} +z {\cal J}_{\rm ex} \right)
\left(J_{\mu}\left\langle J_{\mu}\right\rangle-\frac{1}{2}\left\langle J_{\mu}\right\rangle^2\right) .
\label{eq:Hmf}
\end{eqnarray}
with $z=6$ the number of nearest neighbors.
The last term in Eq.~(\ref{eq:Hmf}), 
$- \frac{1}{2} \left\langle J_{\mu}\right\rangle ^{2}$, has 
no effect on the calculated thermal expectation values of
the $\hat x$ and $\hat z$ components of the magnetization,
and can be dropped. 
The off-diagonal terms, $\mathcal{L}_{\mu\nu}$ with $\mu\ne \nu$,
vanish due to the lattice symmetry.
We employ the Ewald technique~\cite{Ewald,Ziman,Leeuw,Born,Melko-JPC} 
to calculate the dipole-dipole interaction, $L_{ij}^{\mu\nu}$, of Eq.~(\ref{eq:Lij}).
By summing over all sites $j$ coupled to an arbitrary site $i$, 
we obtain the coefficients $L_{\mu\nu}$ listed in Table \ref{tab:Ls}. 
The considered Ewald sums ignore a demagnetization term~\cite{Melko-JPC} 
and  our calculations can therefore be interpreted as corresponding 
to a long needle-shape sample.

\begin{table}[ht]
\begin{tabular}{|c|c|c|c|}
\hline 
&
$L_{xx}$&
$L_{yy}$&
$L_{zz}$\tabularnewline
\hline
\hline 
Tb(OH)$_{3}$&
-11.43&
-11.43&
-28.01\tabularnewline
\hline 
Dy(OH)$_{3}$&
-11.40&
-11.41&
-28.20\tabularnewline
\hline 
Ho(OH)$_{3}$&
-11.38&
-11.37&
-28.45\tabularnewline
\hline
\end{tabular}

\caption{Dimensionless lattice sums calculated with the values of $c/a$ taken
from Table \ref{tab:Latice-const}. \label{tab:Ls}}
\end{table}

We diagonalize numerically ${\cal H}_{\rm MF}$ in Eq.~(\ref{eq:Hmf}), 
and calculate self-consistently the thermal averages of $J_{x}$ and $J_{z}$ operators,
from the expression
\begin{equation}
\left\langle J_{\mu}\right\rangle =
\frac{{\rm Trace}[J_{\mu}\exp(-\mathcal{H}_{\rm MF}/T)]}
{{\rm Trace}[\exp(-\mathcal{H}_{\rm MF}/T)]},
\label{eq:Jz}
\end{equation}
where $\mu$ stands for $x$ and $z$. $\left\langle J_{y}\right\rangle = 0$
due to the lattice mirror symmetries and since $\bm{B}$ is applied along $\hat{x}$.

For a given $B_x$, we find the value of the critical temperature,
$T_c(B_x)$, at which the order parameter, $\langle J_z\rangle$, vanishes.
The resulting $B_x$ vs $T$ phase diagrams, obtained that way, 
using {\it all} sets of CFP from Table \ref{tab:CFP},
are shown in Fig. \ref{fig:Phase-diagrams}.
In the main panel, we plot the phase diagrams for Ho(OH)$_3$ and Dy(OH)$_3$, 
using Scott {\it et al.}'s CFP~\cite{Scott-thesis,Scott-Tb-2,Scott-Dy,Scott-Tb}.
The top inset shows the $B_x$ vs $T$ phase diagrams for Tb(OH)$_3$, 
using two available sets of CFP.
This indicates that, for Tb(OH)$_3$, the critical field $B_x(T)$  reaches very quickly 
the upper limit of magnetic fields attainable with commercial magnets.
The bottom inset shows the $B_x$ vs $T$ phase phase diagrams for
Ho(OH)$_3$ and Dy(OH)$_3$ using Karmakar {\it et al.}'s CFP~\cite{Karmakar-Ho,Karmakar-Dy}. 
Although the diagrams differ quantitatively for the two sets of CFP, the overall qualitative
trend is the same for both sets.
Table \ref{tab:TcBx} lists the mean-field estimates of $T_c$ and $B_x^c$ 
together with the experimental
 values of $T_c$ \cite{Catanese-DyHo,Catanese-Tb,Catanese-thesis}.

There are two contributing factors behind the difference between the experimental
and mean-field values of $T_c$ in Table~\ref{tab:TcBx} and,
 presumably once they are experimentally
determined, those for  $B_x^c$.
Firstly, in obtaining those mean-field values from
Eq.~(\ref{eq:Hmf})
and 
Eq.~(\ref{eq:Jz}), we neglected the (presumably) antiferromagnteic nearest-neighbor
exchange ${\cal J}_{\rm ex}$ which would contribute to a depression of
both the critical ferromagnetic temperature $T_c$ 
and $B_x^c$. Secondly, mean-field thgeory neglects correlations in the thermal
and quantum fluctuations which would also contribute to reduce
$T_c$ and $B_x$. 
From the comparison of mean-field theory~\cite{Chakraborty}
and quantum Monte Carlo~\cite{Chakraborty,Tabei-MC} for LiHoF$_4$, we would anticipate
that our mean-field estimates of $T_c$ and $B_x$ are accurate within 20\% to 40\%,
notwithstanding the uncertainty on the crystal field parameters.

By seeking a self-consistent solution for 
$\langle J_z\rangle$, starting from either the
fully polarized or weakly 
polarized state, two branches of solutions are obtained at
low temperature and large $B_x$ for Dy(OH)$_3$.
This suggests a first order PM to FM transition 
when using either set of CFP for this material.
This result was confirmed by a more thorough
investigation (see Section V below). 
The top right inset of Fig. \ref{fig:Tricrital} shows the
behavior of the 
$\langle J_z\rangle$ as a function of $B_x$ for $T=0.3$ K,
illustrating the transition field and the limits for the 
superheating and supercooling regime. 
The black dot in the main panel and inset
of Fig. \ref{fig:Phase-diagrams} shows the
location of the tricritical point (see Section V).
Note that the $B_x$ value at the tricritical point is $\sim$ 4.85 T
using the CFP of Scott {\it et al.} 
(main panel of Fig. \ref{fig:Phase-diagrams})~\cite{Scott-thesis,Scott-Tb-2,Scott-Dy,Scott-Tb}.  
Hence, the occurrence of a first order transition here is not directly
connected to the degeneracy occurring between the two lowest energy
levels at $B_x=3.92$ T using the same set of CFP 
(see inset of Fig. \ref{fig:EnergyLevels} for Dy(OH)$_3$).
A zoom on the low temperature regime and the vicinity of the tricritical
point for Dy(OH)$_{3}$ is shown in Fig. \ref{fig:Tricrital}.
The calculation details needed to obtain the phase diagram 
of Fig. \ref{fig:Tricrital} are described in Section \ref{sec:G-L}.
The existence of a first order transition at strong $B_x$ in Dy(OH)$_3$ depends
on the details of the chosen Hamiltonian in Eq.~(\ref{eq:Hfull}).
For example,  as discussed in Section \ref{sec:1stOrdBzJex},
a sufficiently strong nearest-neighbor antiferromagnetic exchange,
${\cal J}_{\rm ex}$, eliminates the first order transition. 
We also discuss in Section \ref{sec:1stOrdBzJex} the role of a longitudinal 
field $B_z$ (along $c$ axis) on the first order transition.
At this time, one must await experimental results to ascertain the specific
low temperature behavior that is at play for strong $B_x$ in Dy(OH)$_3$.

\begin{table}
\begin{tabular}{|c|c|c|c|}
\hline
{\small Crystal}&
{\small experimental} $T_{c}$ {[}K]&
MFT $T_{c}$ {[}K]&
{\small MFT} $B_{x}^{c}$ {[}T]\tabularnewline
\hline
\hline
{\small Ho(OH)$_{3}$}&
2.54&
4.28&
4.35\tabularnewline
\hline
{\small Dy(OH)$_{3}$}&
3.48&
5.31&
5.03\tabularnewline
\hline
{\small Tb(OH)$_{3}$}&
3.72&
5.59&
54.81\tabularnewline
\hline
\end{tabular}

\caption{\label{tab:TcBx}Experimental values of critical 
temperatures $T_{c}$~\cite{Catanese-DyHo,Catanese-Tb}
and mean-field theory (MFT) estimates for $T_{c}$ and $B_{x}^{c}$.}
\end{table}

\begin{figure}[ht]
\includegraphics[width=1.0\columnwidth,keepaspectratio]{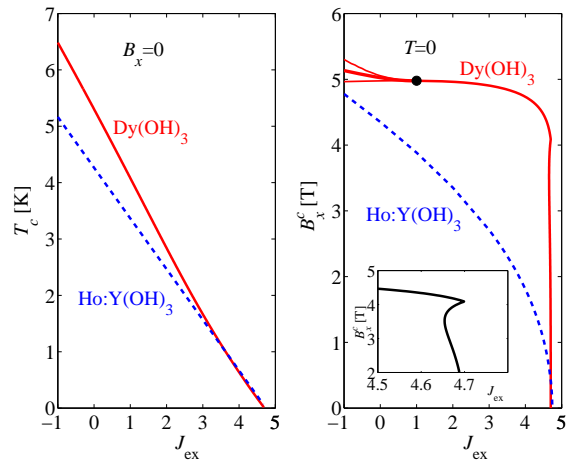}
\caption{The effect of the value of the exchange constant 
on the phase boundary, $T_c$ vs $J_{\rm ex}$ at $B_x$=0 (left) and $B_x^c$ vs $J_{\rm ex}$ at $T=0$ (right).
Solid and dashed lines are used for Dy(OH)$_3$ and Ho(OH)$_3$, respectively. 
The inset of the right graph shows a focus on the features of the high $J_{\rm ex}$ regime in the Dy(OH)$_3$ plot.
The dot in the right panel indicates the tricritical point. 
The two additional lines at the left side of the tricritical point mark the 
limits of superheating and supercooling regimes. 
In the calculations, the CFP of Ref.~[\onlinecite{Scott-thesis}] were used. 
\label{fig:TcBx_vs_Jex}}
\end{figure}

We now briefly analyze
the effect of a nonzero exchange interaction.
The dependence of  the critical temperature, $T_c$, 
and the critical transverse field, $B_x^c$, 
on the exchange constant, $J_{\rm ex}$,
is plotted in Fig. \ref{fig:TcBx_vs_Jex}.
The dot on the $B_x$ vs $J_{\rm ex}$ plot for Dy(OH)$_3$ 
indicates the threshold value of $J_{\rm ex}$,
$J_{\rm ex}^{\rm 2nd}$=0.995, above which the first order transition ceases to exist.
The dependence of the existence of the first order transition  on $J_{\rm ex}$
is discussed in some detail in Section \ref{sec:1stOrdBzJex}.
For $J_{\rm ex} < J_{\rm ex}^{{\rm 2nd}}$, the thinner lines correspond to the 
boundary of the supercooling and superheating regime. 
In the mean-field theory presented here, $J_{\rm ex}$ 
simply adds to the interaction constant
$\mathcal{L}_{\mu\mu}$ with $\mu=x,y,z$
in Eq.~(\ref{eq:Hmf}) (see Table ~\ref{tab:Ls}).
Hence, beyond a threshold value of  $J_{\rm ex}$, 
the system no longer admits a long range ordered ferromagnetic phase.
In the case of Dy(OH)$_{3}$, B$_{x}^{c}$ stays almost unchanged as 
a function of $J_{{\rm ex}}$, until it drops very sharply when 
$\mathcal{L}_{zz} + z\mathcal{J}_{\rm ex}=0$ ($z=6$).
In the inset of Fig. \ref{fig:TcBx_vs_Jex}, we focus on the regime 
where $B_{x}^{c}$ vs $J_{{\rm ex}}$ plot sharply drops. 
The cusp at $B_x=$3.92 T is a consequence of the degeneracy of the 
lowest energy eigenstates (see Fig. \ref{fig:EnergyLevels}). 
As will be shown in detail in the next section for the
Ho(OH)$_3$ system, 
the energy gap separating transverse-field-splitted levels 
of the ground state doublet plays the role of 
an effective transverse field $\Gamma(B_x)$ acting on effective Ising spins. 

\section{Effective $S=1/2$ Hamiltonian \label{Heff} \label{sec:Ising}}

In this section we show that Ho(OH)$_{3}$ and Tb(OH)$_{3}$
can be described with good accuracy by an effective TFIM Hamiltonian.
On the other hand, although Dy(OH)$_{3}$ has been referred in
the literature as an Ising material \cite{Catanese-DyHo,Wolf-Tigges},
we find that it is not possible to describe the magnetic properties of this material
within the framework of an effective Ising Hamiltonian that neglects the effect
of the excited crystal field states.

To be able to identify a material as a realization of an effective 
{\it  microscopic} Ising model,
the following conditions should apply \cite{Wolf}:

\begin{itemize}
\item There has to be a ground state doublet or a close pair of singlets
that are separated from the next energy level by an energy gap that
is large in comparison with the critical temperature. This ensures that at the
temperatures of interest only the two lowest levels are significantly
populated. 

\item To first order, there has to be no transverse susceptibility. It
means that there should be no matrix elements of $(J_{x},J_{y})$ operators
between the two states of the ground doublet.

\item Furthermore, the
longitudinal (in the easy axis direction) susceptibility has to be
predominantly controlled by the two lowest levels. 
In other words, there has
to be no significant mixing of the states of the lowest doublet
with the higher levels via the internal mean field along the Ising direction. 
In more technical terms, the van Vleck susceptibility should play a
negligible role to the non-interacting (free ion) susceptibility near
the critical temperature~\cite{Jensen}.

\item 
In setting up the above conditions, one is in effect 
requesting that a material be describable as a TFIM 
from a miscroscopic point of view.
However, one can, alternatively,
ask whether the quantum critical point of a given material 
is in the same universality class as the relevant 
transverse field Ising model. In such a case,
as long as transition is second order, 
then sufficiently close to the quantum critical point, 
a mapping to an effective TFIM is always in principle possible.
However,  it can be difficult to estimate the pertinent 
parameters entering
the Ginzburg-Landau-Wilson theory describing the transition.
From that perspective, the first and the third 
point above are always fulfilled sufficiently close
to a second order quantum critical point~\cite{referee}.

\end{itemize}

The first condition is not satisfied in the case of Dy(OH)$_{3}$.
The energy gap of 7.8 cm$^{-1}$ $\approx$ 11.2 K
is not much larger than the mean-field 
critical temperature $T_{c} \sim 5.31~{\rm K}$.
Hence, at temperatures close to $T_{c}$, the first excited
doublet state is also significantly populated. 
Furthermore, and most importantly, 
in the context of a field-induced quantum phase transition, 
the third condition above is also not satisfied. 
Hence, even at low temperatures,  
because of the admixing of the two lowest energy 
states with the higher energy levels that is induced
via the internal (mean) field from the surrounding ions, 
Dy(OH)$_3$ cannot by described by an effective 
{\it microscopic} 
Ising model that solely considers the ground 
doublet and ignores the excited crystal field states.
This effect and the associated role 
of nonzero $J_{z}$ matrix elements between the ground state 
and higher crystal field levels
is discussed in more detail in  Appendix \ref{sec:Perturbative}.
As an interesting consequence of this participation of the higher energy levels,
we predict that, unlike in the TFIM of
Eq.~(1), a first order phase transition
may occur at high transverse field in
Dy(OH)$_3$ (see Section \ref{sec:G-L}).

For Ho(OH)$_3$ and Tb(OH)$_3$ we construct an effective Ising Hamiltonian,
following the method 
of Refs.~[\onlinecite{Chakraborty,Tabei-MC,TabeiVernay}].
We diagonalize
exactly the noninteracting Hamiltonian, ${\cal H}_0$ of Eq.~(\ref{eq:HZeeman}), 
for each value of the transverse field, $B_{x}$. 
We denote the two lowest states
by $\ket{\alpha(B_{x})}$ and $\ket{\beta(B_{x})}$ and their energies
by $E_{\alpha}(B_{x})$ and $E_{\beta}(B_{x})$, respectively. 
a transverse field enforces a unique choice of basis, 
in which the states can be interpreted as 
$\ket{\rightarrow}$ and $\ket{\leftarrow}$ in the Ising subspace.
We introduce a new 
$\ket{\uparrow}$ and $\ket{\downarrow}$ basis, in which the
$J_{z}$ matrix elements are diagonal,
by performing a rotation 
\begin{equation}
\begin{array}{l}
\ket{\uparrow}=\frac{1}{\sqrt{2}}(\ket{\alpha(B_{x})}+\exp(i\theta)\ket{\beta(B_{x})}),\\
\ket{\downarrow}=\frac{1}{\sqrt{2}}(\ket{\alpha(B_{x})}-\exp(i\theta)\ket{\beta(B_{x})}).
\end{array}
\label{eq:rot}
\end{equation}
In this basis, the effective single ion Hamiltonian, describing the two lowest states, is of the form
\begin{equation}
\mathcal{H}_{T}=E_{{\rm CM}}(B_{x})-\frac{1}{2}\Delta(B_{x})\sigma^{x},
\label{eq:TFHam}
\end{equation}
where $E_{{\rm CM}}(B_{x})=\frac{1}{2}(E_{\alpha}(B_{x})+E_{\beta}(B_{x}))$
and $\Delta(B_{x})=E_{\beta}(B_{x})-E_{\alpha}(B_{x})$. 
Thus the
splitting of the ground state doublet plays the role of a transverse
magnetic field, $\Gamma \equiv \frac{1}{2}\Delta(B_x)$ in Eq.~(1).
In the case of Tb(OH)$_{3}$, after performing the rotation
(\ref{eq:rot}), even at $B_{x}=0$, a small transverse field term
($\Gamma=\frac{1}{2}\Delta(0)>0$) is present in Hamiltonian (\ref{eq:TFHam}). 
For Dy(OH)$_3$ and Ho(OH)$_3$, the
splitting of the energy levels, obtained via exact diagonalization 
was already discussed at the end of Section \ref{Materials} and   
is shown in Fig. \ref{fig:EnergyLevels}.
To include the interaction terms in our Ising Hamiltonian, we expand
the matrix elements of $J_{x}$, $J_{y}$ and $J_{z}$ operators in
terms of the
$\sigma^\nu$ ($\nu=x,y,z$) Pauli matrices and a unit matrix, $\sigma^0\equiv {\openone }$,
\begin{equation}
J_{i,\mu} = C_{\mu}{\openone }
+\sum_{\nu=x,y,z}C_{\mu\nu}(B_{x})\sigma_i^{\nu}.
\label{eq:pauli}
\end{equation}
 %
\begin{figure}[ht]
\includegraphics[width=1.0\columnwidth,keepaspectratio]{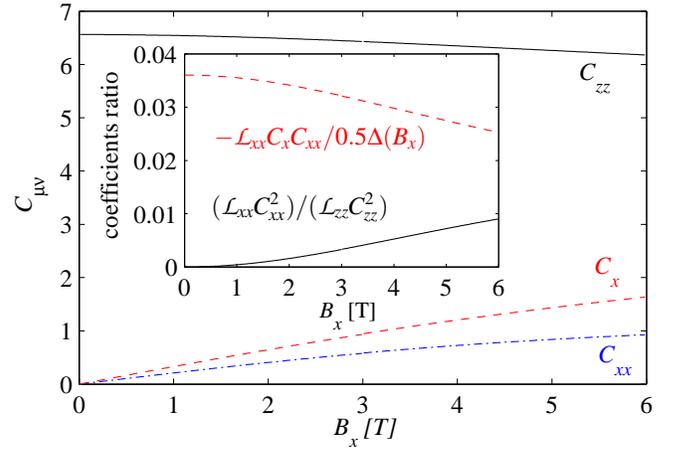}

\caption{The coefficients in the projection of ${\bm J}$ 
operators onto the two dimensional Ising subspace for Ho(OH)$_{3}$. The inset shows
the ratios of the coefficients present in Eq.~(\ref{eq:IsHamLong}).
\label{fig:C}}
\end{figure}

By replacing all $J_{i,{\mu}}$ operators in the interaction term of 
Hamiltonian (\ref{eq:Hfull}) by the two dimensional representation of 
Eq.~(\ref{eq:pauli}), one obtains in general a lengthy Hamiltonian
containing all possible combinations of spin-$\frac{1}{2}$ interactions.
In the present case, the resulting
Hamiltonian is considerably simplified by the crystal symmetries and
the consequential vanishing of off-diagonal elements of the 
interaction matrix $\mathcal{L}_{\mu\nu}$.
This would not be the case for diluted Ho$_x$Y$_{1-x}$F$_4$ 
(see Ref.~[\onlinecite{TabeiVernay}]]).
After performing the transformation in Eq.~(\ref{eq:rot}), we have 
$J_i^{z}=C_{zz}\sigma_i^{z}$,
$J_i^{y}=C_{yy}\sigma_i^{y}$
and $J_i^{x}=C_{xx}\sigma_i^{x}+C_{x}{\openone }$.
Hence, we can rewrite the mean-field Hamiltonian (\ref{eq:Hmf}) in the form 
\begin{widetext}
\begin{eqnarray}
\mathcal{H}_{\rm MF} & = & 
(\mathcal{L}_{zz} + z{\cal J}_{\rm ex} ) C_{zz}^{2} m_z\sigma^{z} +
\left(\mathcal{L}_{xx}C_{x}C_{xx} - \frac{1}{2}\Delta(B_{x})\right)\sigma^x \nonumber \\ 
& + & 
(\mathcal{L}_{xx} + z{\cal J}_{\rm ex}) C_{xx}^{2}  m_x\sigma^{x}
+
(\mathcal{L}_{yy} + z{\cal J}_{\rm ex}) C_{yy}^{2} m_y\sigma^{y},
\label{eq:IsHamLong}
\end{eqnarray}
\end{widetext}

where $m_{\nu}\equiv\left\langle \sigma^{\nu}\right\rangle$ 
and $\left\langle \ldots \right\rangle$ denotes a Boltzmann thermal average.

The $C_{zz}$, $C_{xx}$ and $C_{x}$ coefficients for Ho(OH)$_{3}$
are plotted in Fig. \ref{fig:C}. 
The inset shows a comparison of the terms in $\mathcal{H}_{\rm MF}$. 
In Ho(OH)$_{3}$, the coefficient ${\it \mathcal{L}_{xx}}{\it C_{xx}}^{2}$ 
(the fourth therm of  $\mathcal{H}_{\rm MF}$) does not exceed 1.5\% of 
the effective transverse field, 
$\Gamma={\it \mathcal{L}_{xx}}{\it C_{x}}{\it C_{xx}}-\frac{1}{2}\Delta(B_{x})$. 
In Tb(OH)$_{3}$, this ratio is even smaller, and we thus neglect it,
further motivated by the fact that doing so decouples $m_z$ from $m_x$ and make 
the problem simpler.
The term 
$\mathcal{L}_{yy} {C_{yy}}^{2}m_y\sigma^{y}$
in Eq.~\ref{eq:IsHamLong} can be omitted, since due to symmetry 
$m_y \equiv \langle \sigma^y\rangle=0$.
The interaction correction, ${\it \mathcal{L}_{xx}}{\it C_{x}}{\it C_{xx}}$, 
to the effective transverse field, $\Gamma$, is of order of 3\% of
$\Gamma$ and we retain it in our calculations.
Thus, we finally write
\begin{equation}
\mathcal{H}_{\rm MF}=P\sigma^{z}+\Gamma\sigma^{x},
\label{eq:IsHam}\end{equation}
where $P={\it \mathcal{L}_{zz}}{\it C_{zz}}^{2}m_{z}$ and 
$\Gamma={\it \mathcal{L}_{xx}}{\it C_{x}}{\it C_{xx}}-\frac{1}{2}\Delta(B_{x})$.

Diagonalizing the Hamiltonian (\ref{eq:IsHam}) 
allows us to evaluate $m_z$ and $m_x\equiv \langle \sigma^x\rangle$,
giving well known formulae~\cite{Sen}:
\begin{equation}
\begin{array}{c}
m_{x}=\frac{\Gamma}{\sqrt{P^{2}+\Gamma^{2}}}\tanh(\sqrt{P^{2}+\Gamma^{2}}/T),\\
m_{z}=\frac{P}{\sqrt{P^{2}+\Gamma^{2}}}\tanh(\sqrt{P^{2}+\Gamma^{2}}/T)\end{array},\end{equation}
 and the phase boundary,
\begin{equation}
T_c(B_x)=\frac{\Gamma(B_{x})}{{\rm atanh}(\frac{\Gamma(B_{x})}{\mathcal{L}_{zz}C_{zz}^{2}})} .
\label{eq:IsPhaseBound}
\end{equation}

\begin{figure}
\includegraphics[width=1.0\columnwidth,keepaspectratio]{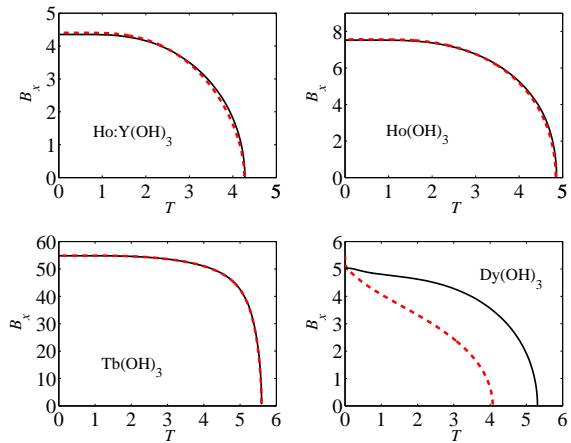}
\caption{Comparison of the phase diagrams obtained with diagonalization of
the full manifold (black, solid lines) and with effective \mbox{spin-1/2} 
Hamiltonian (red, dashed lines).
Calculation for Ho(OH)$_3$ were performed using the CFP 
of Karmakar {\it et al.}~\cite{Karmakar-Ho,Karmakar-Dy}.
\label{fig:Comparison}}
\end{figure}

In Fig. \ref{fig:Comparison}, we show 
that Eq.~(\ref{eq:IsPhaseBound}) yields a phase diagram that only
insignificantly differs from the one obtained from the full 
diagonalization of
${\cal H}_{\rm MF}$ in Eq.~(\ref{eq:Hmf})
shown in Fig. \ref{fig:Phase-diagrams},
in the case of Ho(OH)$_{3}$, and, in the case of Tb(OH)$_{3}$, 
the discrepancy is even smaller because the energy gap to the third crystal field state, 118 cm$^{-1}\simeq$ 170 K,
is very large compared to $T_c^{\rm MF}=5.59$ K.

As alluded to above, in the case of Dy(OH)$_{3}$,
a description in terms of an 
effective Ising Hamiltonian method does
not work because of the admixing 
between states of the two lowest doublets
induced by the local mean-field that is
proportional to $\Jz $ (see Appendix A).
The dashed line in the last panel of Fig. \ref{fig:Comparison} shows the incorrect phase
diagram obtained for Dy(OH)$_{3}$ obtained using
an effective spin-$1/2$
Hamiltonian constructred from only the ground doublet.
It turns out that a form of the method of Section \ref{sec:Ising} can still
be used.
However,
instead of keeping only two levels in the interaction Hamiltonian,
one needs to retain at least four states. 
In analogy with the procedure in 
Section \ref{sec:Ising},  we diagonalize
the single ion Hamiltonian ${\cal H}_0$  
of Eq. (\ref{eq:HZeeman}) which consist of the crystal
field Hamiltonian and the transverse field term. Next, we write an effective interaction
Hamiltonian using the four (or six) lowest eigenstates of ${\cal H}_0$. 
The resulting effective Hamiltonian is then 
used in the self-consistent Eqs. (\ref{eq:Jz}). 
For example, for $B_x=4.8$ T, proceeding by keeping only the
four lowest eigenstates of ${\cal H}_0$ to construct
the effective Hamiltonian,  one finds a critical temperature 
that is only about 3\% off compared to a calculation that keeps all
16 eigentates of ${\cal H}_0$. This difference drops below 1\% when
keeping the 6 lowest eigenstates of ${\cal H}_0$.

Having explored the quantitative validity of the \mbox{spin-1/2} TFIM 
description of Ho(OH)$_3$ and Tb(OH)$_3$ in nonzero $B_x$,   
we now turn to the problem of the first order PM to FM transition
at large $B_x$ and low temperature in Dy(OH)$_3$,
exposed in the numerical solution of the self-consistent equations
comprised in Eq.~(\ref{eq:Jz}) (with $\mu=x,z$).

\section{First order transition}

The first order transition in Dy(OH)$_{3}$ takes its origin
in the sizeable 
admixing among the four lowest levels induced by the
the local mean-field
that is proportional to $\Jz$. 
Under the right temperature and field conditions,
two free-energy equivalent configurations  
can exist: an ordered state with some not infinitesimally
small magnetization, $\Jz>0$, and a state with zero magnetic moment, $\Jz=0$. 
To simplify the argument, we consider how this occurs at $T=0$. 
At first, let us look at the situation when the longitudinal internal
mean field induces an admixing
 of the ground state with the first excited state only (as in the TFIM). 
In such a case, there is only a
quadratic dependence of the ground state energy 
on  the longitudinal mean field,
$B_z^{\rm MF}$, and, consequently, only one energy minimum is possible.
 Now, if there is an
admixing of the ground state and at least three higher levels,
 the dependency of 
the ground state energy on  $B_{z}^{{\rm MF}}$  is of fourth order and two energy 
minima are, in principle, possible. 
Thus, at a certain value of external
 parameters the system can acquire two energetically equivalent states, 
one with zero and the other with a non-zero magnetization. 
When passing through this point, either by varying the transverse field or the temperature, 
a first order phase transition characterized by a magnetization discontinuity occurs.
To make this discussion more formal, we now proceed with a construction
of the Ginzburg-Landau theory for Dy(OH)$_3$ for arbitrary $B_x$ in the regime
of $B_x$ and $T$ values where the paramagnetic to ferromagnetic transition is
second order.
This allows us to determine the the tricritical transverse field value
above which the transition becomes first order.



\subsection{Ginzburg-Landau Theory\label{sec:G-L}}

To locate the tricritical point for Dy(OH)$_3$, 
we perform a Landau expansion of the
mean-field free energy,
${\cal F}_{\rm MF}(\Jx ,\Jz )$.
Next, we minimize ${\cal F}_{\rm MF}$ with
respect to $\Jx $,
leaving $\Jz $ as the only free parameter.
The mean-field free energy can be written in the form 
\begin{eqnarray}
\mathcal{F}_{\rm MF} 
\left(\Jx ,
\Jz \right) 
& = & -T\log Z\left(\Jx ,\Jz \right)
\nonumber \\
 & - & \frac{1}{2}\left(\mathcal{L}_{xx}\Jx ^{2}+\mathcal{L}_{zz}
\Jz ^{2}\right),
\label{eq:F}
\end{eqnarray}
where $Z\left( \Jx, \Jz \right)$ is the partition function. 

Just below the transition, 
in the part of the phase diagram where the transition is second order, 
$\Jz $
is a small parameter (i.e. has a small dimensionless numerical value). 
We therefore make an expansion for $\Jx $
as a function of $\Jz $, which we  write
it in the form:
\begin{equation}
\Jx =
\Jx _{0}+\delta ({\Jz}).
\label{eq:Jx_exp}
\end{equation}
$\Jx _{0}$ 
is the value of $\Jx$ that
extremizes ${\cal F}_{\rm MF}$ 
when $\Jz =0$.
$\delta({\langle  J_z \rangle })$
is a perturbatively small function of 
$\Jz $,
which we henceforth simply denote $\delta$,  
and which is our series expansion small parameter for $\langle J_x \rangle$.
Substituting expression (\ref{eq:Jx_exp}) to $\mathcal{H}_{\rm MF}$ of 
Eq.~(\ref{eq:Hmf}), and setting $J_{\rm ex}=0$ for the time being, we have 
\begin{eqnarray}
\mathcal{H} & = & {\cal H}_{\rm cf}({\bm J}_{i})-g\mu_{\rm B}B_{x}J_{x} \nonumber  \\
 & + & \mathcal{L}_{xx}J_{x}(\Jx _{0}+\delta)+\mathcal{L}_{zz}J_{z}
\Jz ,
\end{eqnarray}
or
\begin{equation}
\mathcal{H} = \mathcal{H}_{0}(B_{x},\Jx _{0})
  +  \mathcal{L}_{xx}J_{x}\delta+\mathcal{L}_{zz}J_{z}\Jz,
\label{eq:Hamil}
\end{equation}
where for brevity, as in Eq.~(\ref{eq:Hmf}), 
the constant term has been dropped because, again, it does not affect
the expectation values needed for the calculation.

The power series expansion of the partition function, and then of
the free energy (\ref{eq:F}), can be calculated from the eigenvalues
of Hamiltonian (\ref{eq:Hamil}). 
Instead of applying standard quantum-mechanical
perturbation methods to Eq.~(\ref{eq:Hamil}), we obtain the
expansion of energy levels as a perturbative, `seminumerical', solution to the 
characteristic polynomial equation
\begin{equation}
\det\left[\mathcal{H}_{0}+\mathcal{L}_{xx}J_{x}\delta+
\mathcal{L}_{zz}J_{z}\Jz -E_{n}\right]=0.
\label{eq:polyn}
\end{equation}
We can easily implement this procedure by using
a computer algebra method
(e.g. Maple$\texttrademark$ or Mathematica$\texttrademark$).
To proceed, we 
substitute a formal power series expansion of the solution 
\begin{equation}
E_{n}=E_{n}^{(0,0)}+E_{n}^{(0,1)}\delta+
E_{n}^{(2,0)}\Jz ^{2}
+
E_{n}^{(2,1)}\Jz ^{2}\delta+\ldots,
\label{eq:expa}
\end{equation}
to Eq.~(\ref{eq:polyn}), containing all the terms of the form 
$E_{n}^{(\alpha,\beta)}\Jz^{\alpha}\delta^{\beta}$, where $\alpha+2\beta \le 6$,
as will be justified below Eq.~(\ref{eq:delta}).
To impose consistency  of the resulting equation obtained 
from Eq.~(\ref{eq:polyn}) and Eq.~(\ref{eq:expa}),
up to sixth order of the expansion in $\Jz$ in),
we need to equate to zero all the coefficient with the required order of 
$\Jz$ and $\delta$, i.e. $\alpha+2\beta \le 6$.
This gives a system of equations that can be numerically solved 
for the coefficients $E_{n}^{(k,l)}$, where $k,l > 0$. 
By $E_{n}^{(0,0)}$ we denote the
eigenvalues of the Hamiltonian $\mathcal{H}_{0}(B_{x},\Jx _{0})$.

We use the perturbed energies, $E_n$, of Eq.~(\ref{eq:expa}) to calculate the partition function
\begin{equation}
Z(\delta,\Jz) = \sum_n e^{-E_n/T}
\label{eq:justZ}
\end{equation}
and substitute it in Eq.~(\ref{eq:F}).
We Taylor expand the resulting expression to obtain the numerical values of 
the expansion coefficients  in the form 
\begin{equation}
{\cal F}_{\rm MF}
=A^{(0,0)}+A^{(2,0)}\Jz ^{2}+A^{(0,1)}\delta+A^{(2,1)}\left\langle J_{z}
\right\rangle ^{2}\delta+\ldots.
\label{eq:FB}
\end{equation}
The free energy 
${\cal F}_{\rm MF}$ 
is a symmetric function of $\Jz $,
so the expansion (\ref{eq:FB}) contains only even powers of $\Jz $.
We minimize ${\cal F}_{\rm MF}$ in Eq.~(\ref{eq:FB}) with respect to $\delta$. 
To achieve this, we have to solve a high order polynomial equation  
$d{\cal F}_{\rm MF}/d\delta=0$. 
Again, we do it by substituting to the equation a formal power series solution
\begin{equation}
\delta({\langle J_z \rangle})
=D_{2}\Jz ^{2}+D_{4}\Jz ^{4}
+\ldots
\label{eq:delta}
\end{equation}
and then solve it for the values of the expansion parameters $D_n$.
Due to symmetry, only even powers of $\Jz $
are present and, from the definition of $\delta$, the constant 
$\langle J_z\rangle$-independent term
is equal to zero. From the form of the
expansion in Eq.~(\ref{eq:delta}), we see
that to finally obtain the free energy expansion in powers 
of $\Jz $,
up to $n$-th order, we need to consider only the terms
 $\Jz ^{\alpha}\delta^{\beta}$
where $\alpha+2\beta\le n$. 
Finally, by substituting $\delta$ from Eq.~(\ref{eq:delta})
into Eq.~(\ref{eq:FB}), we obtain the power series expansion of the free energy
in the form:
\begin{equation}
{\cal F}_{\rm MF}=
C_{0}+C_{2}\Jz ^{2}+C_{4}\Jz ^{4}+
C_{6}\Jz ^{6}.
\label{eq:Fexp}
\end{equation}

In the second order transition region, the condition $C_{2}=0$ with
$C_{4}>0$ parametrizes the phase boundary. The equations $C_2=C_{4}=0$
gives the condition for the location of the
tricritical point. In the regime where $C_{4}<0$, the condition
$C_{2}=0$ gives the supercooling limit. The 
first order phase transition boundary is located
where the free energy has the same value at both local minima. 
Increasing the value of the control parameters, $T$ and $B_{x}$,
above the critical value, until the second (nontrivial)
local minimum of ${\cal F}_{\rm MF}$ vanishes,
gives the superheating limit.

\begin{figure}[ht]
\includegraphics[width=1.0\columnwidth,keepaspectratio]{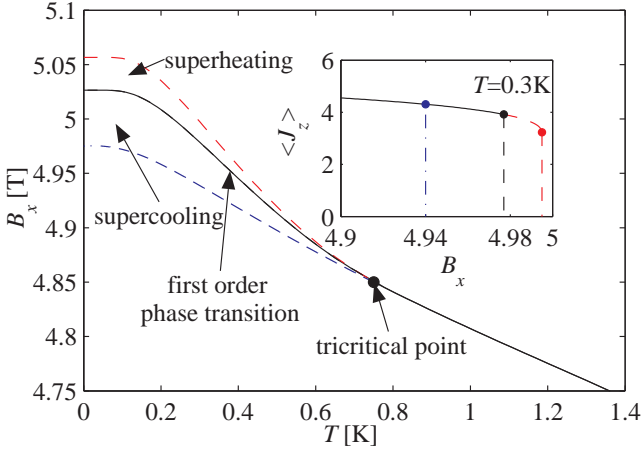}
\caption{Tricritical behavior of Dy(OH)$_{3}$.
The continuous line marks the phase boundary. 
The dot indicates the location of the tricritical point. The upper
 and lower lines (dashed) are the limits of the superheating and
supercooling regimes, respectively. 
As an example, the inset shows the order parameter $\Jz $
vs $B_{x}$, at the temperature $T=$0.3 K.
For this temperature, the phase transition occurs at $B_x\approx 4.98$ T.
The upper dashed extension of
the solid line  corresponds to the superheating limit. The dash-dotted line
to the left of the tricritical point shows the transition at the supercooling limit. 
\label{fig:Tricrital}}
\end{figure}

\begin{figure}[ht]
\includegraphics[width=1.0\columnwidth,keepaspectratio]{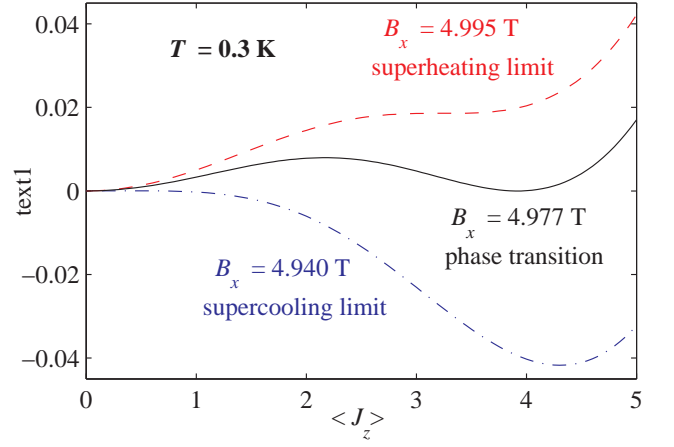}
\caption{Free energy\label{fig:Free-energy} vs average magnetic moment $\Jz $
at the temperature of 0.3 K. 
The free energy at the first order phase transition is plotted with continuous line.
The upper (dashed) line shows the free energy at the boundary of
the superheating limit. 
The lower (dot-dashed) line shows the free energy when the system passes 
through the supercooling limit. 
All the plots show the free energy displaced by a 
constant chosen such that their shape can be compared. }
\end{figure}

The location of the 
tricritical point is  $T_c^{\rm TCP}=$0.75 K, $B_x^{\rm TCP}=$4.85 T. 
We show in Fig. \ref{fig:Tricrital} the first and the second order
transition phase boundary; the tricritical point is marked with a dot.
In the first order transition regime, the superheating and supercooling
limits are also plotted. 
$\langle J_z \rangle $ ceases to be a small parameter for values of
$T$ and $B_x$ `away' from the tricritical point. 
Thus, the two upper curves in the phase diagram of Fig. \ref{fig:Tricrital}
are determined from a numerical search for both local minima of
the exact mean-field free energy in Eq.~(\ref{eq:F}) without
relying on a small $\Jz $ and $\delta(\Jz )$ expansion.
The supercooling limit
is calculated from the series expansion (\ref{eq:Fexp}) and determined by
the condition $C_2=0$.

In the inset of Fig. \ref{fig:Tricrital}, we show
the average magnetic moment, $\Jz$,
as a function of the transverse field, at the temperature of 0.3 K.
The dots and the dashed lines mark the supercooling limit, first order
phase boundary and the superheating limit, in order of increasing $B_x$.
The shape of the free energy at these three characteristic 
values of the magnetic field, $B_x$, at temperature of 0.3 K, 
is shown in Fig. \ref{fig:Free-energy}.

In Fig. \ref{fig:Free-energy}, we plot the free
energy as a function of $\Jz $, where
$\Jx $ is minimizing ${\cal F}_{\rm MF}$ as a function of
$\Jz $ at $T=0.3$ K. 
Free energy at the phase transition ($B_x\approx$ 4.977 T) is plotted 
with a continuous line. The dashed and dot-dashed plots show 
free energy at the superheating and supercooling limits, 
at $B_x\approx$  4.995 T and $B_x \approx$  4.940 T, respectively. 
The free energy clearly shows the characteristic structure (e.g. barrier)
 of a system with a first order transition.
It would be interesting to investigate whether the
real Dy(OH)$_3$ material exhibits such a $B_x-$induced first order PM to
FM transition at strong $B_x$.
In the event that the transition is second order down to $T=0$ and $B_x=B_x^c$,
Dy(OH)$_3$ would offer itself as another material to investigate
transverse field induced quantum criticality
(see 4$^{\rm th}$ item in the list at the beginning of Section IV).
However, a quantitative microscopic description at strong $B_x$ 
would nevertheless require that the contribution
of the lowest pairs of excited crystal field states be taken into account.

One may be tempted to relate the existence of a first order transition
in Dy(OH)$_3$, on the basis of Eq.~(\ref{eq:FB}),
with two expansion parameters $\langle J_z\rangle$ and $\delta$, 
to the familiar problem where a free-energy function, ${\cal F}(m,\epsilon)$,
of two order parameters $m$ and $\epsilon$, 
\begin{eqnarray*}
{\cal F}(m,\epsilon) & = & \frac{a}{2}m^2 + \frac{|b|}{4}m^4 + \frac{|c|}{6}m^6 
+\frac{K}{2}\epsilon^2 - g\epsilon m^2
\end{eqnarray*}
displays a first order transition when $g^2/K > b/2$.
However, we have found that this analogy is not useful and the
mechamism for the first order transition is not trivially 
due to the presence of two expansion parameters, $\Jz$ and $\delta$,
in the expansion (\ref{eq:FB}). 
It is rather the complex specific details of the crystal
field Hamiltonian for Dy(OH)$_3$ that are responsible for the first
order transition. For example, at a qualitative level,
a first order transition still occurs even if $\delta(\Jz )$, 
in Eq.~(\ref{eq:Jx_exp}), is taken to be 0,
for all values of  $\Jz$.

\subsection{The effect of longitudinal magnetic field and exchange
interaction on the existence of first order transition in Dy(OH)$_{3}$\label{sec:1stOrdBzJex}}
\begin{figure}
\includegraphics[width=1.0\columnwidth,keepaspectratio]{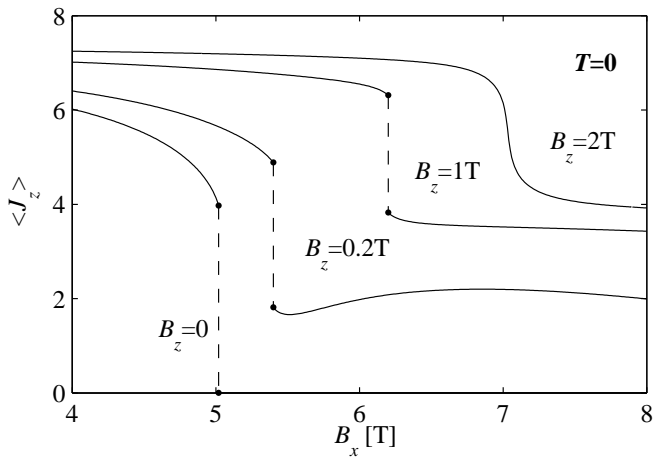}
\caption{\label{fig:JzJxBx}Average magnetic moment, $\Jz$,
vs transverse magnetic field, $B_{x}$, at $T=0$, $J_{\rm ex}=0$, for different
values of longitudinal magnetic field, $B_{z}$.}
\end{figure}
\begin{figure}[ht]
\includegraphics[width=1.0\columnwidth,keepaspectratio]{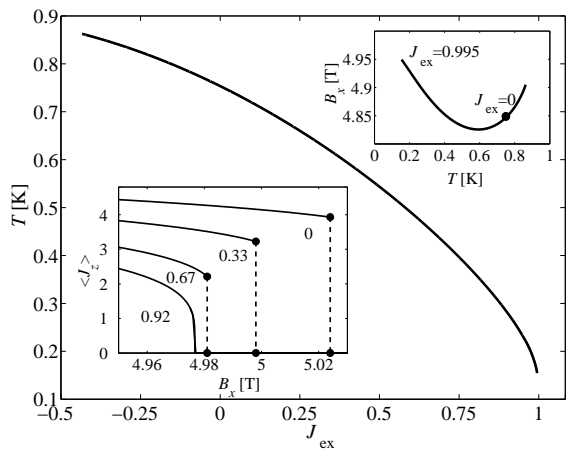}
\caption{Temperature corresponding to the TCP as a function of 
nearest-neighbor  exchange constant $J_{\rm ex}$. 
The upper inset shows position of the TCP in the phase diagram
plane for the same range of exchange constant as in the main plot.
For an exchange $J_{\rm ex} > J_{\rm ex}^{\rm 2nd}$, with
$J_{\rm ex}^{\rm 2nd} = 0.995$,
the tricritical point at $T>0$ ceases to exist. 
The lower inset shows average magnetic moment
$\Jz $ vs transverse field $B_{x}$
at temperature $T=$0, for the shown values of exchange constant $J_{\rm ex}$.}
\label{fig:TCPvsJex}
\end{figure}

Having found that the PM to FM transition may be first order in Dy(OH)$_3$
at large $B_x$ (low $T$), it is of interest to investigate briefly two effects 
of physical relevance on the predicted first order transition. 
Firstly, since the transition is first order from $0\le T\le T^{\rm TCP}$, 
one may ask what is the critical {\it longitudinal} field, $B_z$, required 
to push the tricritical point from finite temperature down to zero temperature.   
Focusing on the CFP of Scott {\it et al.} from 
Refs.~[\onlinecite{Scott-thesis,Scott-Tb-2,Scott-Dy,Scott-Tb}],  
we find that a sufficiently strong
magnetic field, $B_z$, applied along the longitudinal $z$ direction destroys the first
order transition, giving rise to an end critical point.
We plot in Fig. \ref{fig:JzJxBx} the magnetization,
$\Jz$, as a function of
$B_{x}$ for different values of 
$B_z$ at $T=0$. We see that
a critical value of $B_z$ is reached 
between 1 T and 2 T, where the first order transition disappears,
giving rise to an end critical point at $T=0$.
Hence, assuming that the low-temperature $B_x$-driven PM to FM transition 
is indeed first order in Dy(OH)$_3$, the results of 
Fig. (\ref{fig:JzJxBx}) indicate that the critical longitudinal field 
for a quantum critical end point is
easily accessible, using a 
so-called vector magnet (i.e. with tunable 
horizontal, $B_x$,
 and vertical, $B_z$, magnetic fields)~\cite{magnet}. 

It was discussed in Section \ref{Numerics} (Fig. \ref{fig:TcBx_vs_Jex}) that the 
(yet undetermined) 
nearest-neighbor exchange interaction, $J_{{\rm ex}}$, affects the
zero $B_x$ critical temperature, $T_c$, and the zero temperature critical 
transverse field, $B_x^c$.
It is also of interest to explore what is the role of $J_{{\rm ex}}$ on the location 
(temperature and transverse field) of the tricritical point in Dy(OH)$_3$.  

We plot in Fig. \ref{fig:TCPvsJex} the temperature corresponding
to the tricritical point (TCP)
 as a function of antiferromagnetic exchange and, in the
upper inset, the location of the TCP on the phase diagram is presented.
The location of the TCP was calculated using the semi-analytical expansion
described in Section \ref{sec:G-L}. We found that the system ceases
to exhibit a first order transition at nonzero temperature when the value of 
nearest neighbor exchange constant, $J_{{\rm ex}}$, exceeds $J_{\rm ex}^{\rm 2nd}=0.995$.
At $B_x$=0, the critical temperature calculated with the 
value of exchange constant $J_{{\rm ex}}$=0.995 is 4.09 K.
In the lower inset
of Fig. \ref{fig:TCPvsJex} we plot the average magnetic moment, $\Jz$,
as a function of $B_x$ at zero temperature, for different
values of $J_{\rm ex}$. 
The top inset shows a parametric plot of the position of the TCP in the $(T, B_x)$ 
plane as $J_{\rm ex}$ is varied.

\section{Conclusion\label{Conclusion}}

We have presented a simple mean-field theory aimed at motivating an experimental 
study of transverse-field-induced phase transitions in the insulating
rare-earth Ising RE(OH)$_3$ (RE=Dy, Ho)  uniaxial dipolar ferromagnetic materials.

In setting out to perform the above calculations, we were mostly motivated
in identifying a new class of materials as analogous as possible
to LiHo$_x$Y$_{1-x}$F$_4$, where interesting phenomena, both in
zero and nonzero applied transverse field $B_x,$ have been observed. 
In particular, we were interested in finding compounds where 
a systematic comparison between a non-Kramers (e.g. Ho$^{3+}$)
and a Kramers (e.g. Dy$^{3+}$) variant could be investigated.
From our study, we are led to 
suggest that an experimental study of the
Dy$_x$Y$_{1-x}$(OH)$_3$ and
Ho$_x$Y$_{1-x}$(OH)$_3$ materials could bring new pieces of information
on the physics that may be at play in LiHo$_x$Y$_{1-x}$F$_4$
and to ascertain if that physics is unique to LiHo$_x$Y$_{1-x}$F$_4$ 
or if it also arises in other diluted dipolar Ising ferromagnets.

Depending on the details of the
Hamiltonian characterizing Dy(OH)$_3$, 
it may be that a first order transition occurs at low temperature
(large $B_x$), due to 
the admixing between the ground doublet and the low-lying crystal 
field states that is induced by the spin-spin interactions.
For the same reason, we find that  Dy(OH)$_3$ is 
not well described by an effective microscopic
 transverse field Ising model (TFIM).
On the other hand, Ho(OH)$_3$ appears to be very well characterized by
a TFIM and, therefore, constitutes 
a highly analogous variant of LiHoF$_4$.
Tb(OH)$_3$ is also very well described by a TFIM.
Unfortunately, in that case, the critical $B_x$, $B_x^c$, appears
prohibitively large to be accessed via in-house commercial magnets.

We hope that our work will stimulate future systematic experimental investigations of
these materials and, possibly, help shed some light on the rather
interesting problems that
pertain to the fundamental nature of classical and quantum
critical phenomena in disordered dipolar systems and which
have been raised by nearly twenty years of study of 
LiHo$_x$Y$_{1-x}$F$_4$.

\acknowledgements

We thank
P. Chakraborty, 
R. Cone, 
L. Corruccini,
J.-Y. Fortin, 
S. Girvin,
R. Higashinaka, 
R. Hill,
Y. Maeno, 
B. Malkin,
H. Molavian,
V. Oganesyan,
T. Rosenbaum, 
D. Silevitch,
A. Tabei, 
K.-M. Tam,
W. Wolf,  
and 
T. Yavors'kii
for useful discussions.
This work was supported by the NSERC of Canada,
the Canada Research Chair Program (Tier I, M.G),
the Canadian Institute for Advanced Research, 
the Canada Foundation for Innovation and the Ontario Innovation Trust.
M.G. thanks the University of Canterbury (UC) for funding
and the hospitality of
the Department of Physics and Astronomy at UC
where part of this work was done.

\appendix

\section{Perturbative calculation of the phase diagram in Dy(OH)$_{3}$ \label{sec:Perturbative}}

To investigate the role of the $J_{z}$ matrix elements between the
two lowest states and the first excited levels 
on the magnetic behavior of Dy(OH)$_3$,
we calculate the
critical temperature for a second order transition
using second order perturbation theory. This method
is exact in the second order phase transition regime, 
where $B_x$ is less than the tricritical field value, $B_x^{\rm TCP}$
($B_x^{\rm TCP} = 4.85~{\rm T}$ when using the CFP of Scott {\it et al.}). 

For a given value of the transverse field, $B_x$, and the corresponding value 
of average magnetization in transverse direction, $\Jx$,
we consider $\mathcal{L}_{zz}J_{z}\Jz$ term as a perturbation to the
reference mean-field Hamiltonian, 
\begin{equation}
\mathcal{H}_{0}={\cal H}_{\rm cf}(\bm{J}_{i})-g\mu_{\rm B}B_{x}J_{x}+
\mathcal{L}_{xx}J_{x}\Jx ,
\label{eq:H0_ref}
\end{equation}
describing the PM phase at a temperature $T>T_c(B_x)$ as in Eq.~\ref{eq:Hmf}
Here too, we have dropped the constant terms.
 The eigenvalues, $E_{p}$, and eigenstates, $\ket{p}$, of the perturbed
Hamiltonian,
\begin{equation}
\mathcal{H}=\mathcal{H}_{0}+\mathcal{L}_{zz}J_{z}\Jz,
\end{equation}
are written in terms of eigenvalues, $E_{p}^{(0)}$, and eigenstates,
$\ket{p^{(0)}}$, of the unperturbed Hamiltonian, $\mathcal{H}_{0}$, of Eq.~(\ref{eq:H0_ref}),
\begin{equation}
E_{p}=E_{p}^{(0)}+\Jz E_{p}^{(1)}+\Jz ^{2}E_{p}^{(2)},\end{equation}
 \begin{equation}
\vert p \rangle = \vert {p^{(0)}} \rangle +
	\Jz \sum_{k\neq p}c_{p,k}^{(1)} \vert {k^{(0)}} \rangle.
\end{equation}
 The coefficients of the perturbative expansion are given by
\begin{equation}
E_{p}^{(1)} = \mathcal{L}_{zz} J_{pp}^{z},
\end{equation}
\begin{equation}
E_{p}^{(2)} = \sum_{k\neq p}\frac{\mathcal{L}_{zz}^2|J_{kp}^{z}|^{2}}{E_{p}^{(0)}-E_{k}^{(0)}}
\end{equation}
and
\begin{equation}
c_{pk}^{(1)}= \frac{\mathcal{L}_{zz}J_{kp}^{z}}{E_{p}^{(0)} - E_{k}^{(0)}},
\end{equation}
 where $J_{kp}^{z}=\bra{k^{(0)}}J_{z}\ket{p^{(0)}}$ are the matrix
elements of the $J_{z}$ operator in the basis of eigenvectors of the unperturbed Hamiltonian, $\mathcal{H}_{0}$.
The applied magnetic field, $B_{x}$, lifts the degeneracy of the Kramers
doublets, thus we can use the nondegenerate perturbation method. The diagonal elements of the $J_{z}$ operator
vanish, hence, the first order correction to energy vanish, $E_{p}^{(1)}=J_{pp}^{z}=0$.

We calculate the thermal average of the $J_{z}$ operator, 
\begin{equation}
\Jz =\sum_{p}\bra{p}J_{z}\ket{p}e^{-E_{p}/T}/Z,\label{eq:Jz1}
\end{equation}
using the perturbed eigenstates, $\ket{p}$, and eigenvalues, $E_{p}$, 
and where $Z=\sum_{p}e^{-E_{p}/T}.$ Keeping only terms up to third
order in $\Jz $ in the expansion of 
Eq.~(\ref{eq:Jz1}), we find
	\begin{equation}	
		\frac{1}{Z}e^{-E_p/T} = n_p^{(0)}\left(1+K_p\Jz^2\right)
	\end{equation}
and 
	\begin{equation}
		\bra{p}J_{z}\ket{p} = 2\mathcal{L}_{zz}\Jz 
\sum_{k\neq p}\frac{|J_{pk}^z|^2}{E_p^{(0)}-E_k^{(0)}} ,\label{eq:pJzp}
	\end{equation} 
where, for convenience, we write $n_{p}^{(0)}=e^{-E_{p}^{(0)}/T}/Z^{(0)}$,  
$Z^{(0)}=\sum_{p}e^{-E_{p}^{(0)}/T}$ and 
$K_{p} = \frac{1}{T}\left(\sum_{k}n_{k}^{(0)}E_{k}^{(2)}-E_{p}^{(2)}\right)$.
Thus, we can write
\begin{equation}
  \Jz   =  2\Jz
  \sum_{p}n_{p}^{(0)}\left(1+K_{p}\Jz ^{2}\right)
\sum_{k\neq p}\frac{\mathcal{L}_{zz}|J_{pk}^z|^2}{E_p^{(0)}-E_k^{(0)}} ,
\end{equation}
and finally, we get 
\begin{equation}
\Jz ^{2}=\frac{1-2\sum_{p,k\neq p}n_{p}^{(0)}\frac{\mathcal{L}_{zz}|J_{pk}^z|^2}{E_p^{(0)}-E_k^{(0)}}}
		{2{\sum}_{p,k\neq p}n_{p}^{(0)}K_p\frac{\mathcal{L}_{zz}|J_{pk}^z|^2}{E_p^{(0)}-E_k^{(0)}}} .
\label{eq:JzFinal}
\end{equation}
Putting $\Jz =0$ in Eq.~(\ref{eq:JzFinal})
we obtain the condition for the critical temperature $T_{c}$:
\begin{equation}
2\sum_{p,k\neq p}\frac{\mathcal{L}_{zz}|J_{pk}^z|^2}{E_p^{(0)}-E_k^{(0)}}e^{\frac{-E_{p}^{(0)}}{T_c}}=1 .
\label{eq:TcEq}
\end{equation}

In solving Eq.~(\ref{eq:TcEq}) for $T_c$, we have to self-consistently
update the value of $\Jx$ in order to diagonalize
$\mathcal{H}_0$ in Eq.~(\ref{eq:H0_ref}) and to find $E_{p}^{(0)}$. Solving 
Eq.~(\ref{eq:TcEq}) with only the 4 lowest energy eigenstates
(after diagonalizing the full transverse field Hamiltonian of Eq.~(\ref{eq:HZeeman})),
yields a phase diagram that is in good agreement 
for $B_x<B_x^{\rm TCP}$ with the phase boundary obtained 
with all crystal field eigenstates (or equivalently from Eq.~(\ref{eq:Jz})).
 
Estimating the values of the elements of the sum in Eq.~(\ref{eq:TcEq}),
one can see that
the matrix elements of the $J_{z}$ operator, mixing the two lowest
states with the excited states, may bring a substantial correction
to the value of the critical temperature obtained when only 
the two lowest eigenstates are considered.
In the low temperature regime, 
one could omit the matrix elements between the states of
the excited doublet, but we have to keep the matrix elements between
the states of the ground doublet and first excited doublet.
The contribution from the further exited states is quite small
because of the increasing 
value of the energy gap present in the denominator in Eq.~(\ref{eq:TcEq}).

At $T$=0, Eq.~(\ref{eq:pJzp}) 
leads to the equation for the critical transverse field, $B_x^c$:
\begin{equation}
\label{eq:BxcPert}
2\sum_{k\neq 1}\frac{\mathcal{L}_{zz}|J_{1k}^z(B_x^c)|^2}{E_1^{(0)}(B_x^c)-E_k^{(0)}(B_x^c)}=1.
\end{equation}
Again, we see that the matrix elements of $J^z$ operator, 
admixing the ground state with excited levels, have to be considered.
Note that since Eq.~(\ref{eq:BxcPert}) 
pertains to the case of zero temperature, this equation is only valid 
in a regime where the transition is second order
 (i.e. when $J_{{\rm ex}} > J_{{\rm ex}}^{\rm 2nd}$). 


\begin{thebibliography}{86}
\expandafter\ifx\csname natexlab\endcsname\relax\def\natexlab#1{#1}\fi
\expandafter\ifx\csname bibnamefont\endcsname\relax
  \def\bibnamefont#1{#1}\fi
\expandafter\ifx\csname bibfnamefont\endcsname\relax
  \def\bibfnamefont#1{#1}\fi
\expandafter\ifx\csname citenamefont\endcsname\relax
  \def\citenamefont#1{#1}\fi
\expandafter\ifx\csname url\endcsname\relax
  \def\url#1{\texttt{#1}}\fi
\expandafter\ifx\csname urlprefix\endcsname\relax\def\urlprefix{URL }\fi
\providecommand{\bibinfo}[2]{#2}
\providecommand{\eprint}[2][]{\url{#2}}

\bibitem[{\citenamefont{Sondhi et~al.}(1997)\citenamefont{Sondhi, Girvin,
  Carini, and Shahar}}]{Sondhi}
\bibinfo{author}{\bibfnamefont{S.~L.} \bibnamefont{Sondhi}},
  \bibinfo{author}{\bibfnamefont{S.~M.} \bibnamefont{Girvin}},
  \bibinfo{author}{\bibfnamefont{J.~P.} \bibnamefont{Carini}},
  \bibnamefont{and} \bibinfo{author}{\bibfnamefont{D.}~\bibnamefont{Shahar}},
  \bibinfo{journal}{Rev. Mod. Phys.} \textbf{\bibinfo{volume}{69}},
  \bibinfo{pages}{315} (\bibinfo{year}{1997}).

\bibitem[{\citenamefont{Sachdev}(1999)}]{Sachdev}
\bibinfo{author}{\bibfnamefont{S.}~\bibnamefont{Sachdev}},
  \emph{\bibinfo{title}{Quantum Phase Transitions}}
  (\bibinfo{publisher}{Cambridge University Press, Cambridge},
  \bibinfo{year}{1999}).

\bibitem[{\citenamefont{Elliott et~al.}(1970)\citenamefont{Elliott, Pfeuty, and
  Wood}}]{Pfeuty}
\bibinfo{author}{\bibfnamefont{R.~J.} \bibnamefont{Elliott}},
  \bibinfo{author}{\bibfnamefont{P.}~\bibnamefont{Pfeuty}}, \bibnamefont{and}
  \bibinfo{author}{\bibfnamefont{C.}~\bibnamefont{Wood}},
  \bibinfo{journal}{Phys. Rev. Lett.} \textbf{\bibinfo{volume}{25}},
  \bibinfo{pages}{443} (\bibinfo{year}{1970}).

\bibitem[{\citenamefont{Chakrabarti et~al.}(1996)\citenamefont{Chakrabarti,
  Dutta, and Sen}}]{Sen}
\bibinfo{author}{\bibfnamefont{B.~K.} \bibnamefont{Chakrabarti}},
  \bibinfo{author}{\bibfnamefont{A.}~\bibnamefont{Dutta}}, \bibnamefont{and}
  \bibinfo{author}{\bibfnamefont{P.}~\bibnamefont{Sen}},
  \emph{\bibinfo{title}{Quantum Ising Phases and Transitions in Transverse
  Ising Models}} (\bibinfo{publisher}{Spinger-Verlag, Heidelberg},
  \bibinfo{year}{1996}).

\bibitem[{\citenamefont{de~Gennes}(1963)}]{deGennes}
\bibinfo{author}{\bibfnamefont{P.~G.} \bibnamefont{de~Gennes}},
  \bibinfo{journal}{Solid State Commun.} \textbf{\bibinfo{volume}{1}},
  \bibinfo{pages}{132} (\bibinfo{year}{1963}).

\bibitem[{\citenamefont{Chakraborty et~al.}(2004)\citenamefont{Chakraborty,
  Henelius, Kjonsberg, Sandvik, and Girvin}}]{Chakraborty}
\bibinfo{author}{\bibfnamefont{P.~B.} \bibnamefont{Chakraborty}},
  \bibinfo{author}{\bibfnamefont{P.}~\bibnamefont{Henelius}},
  \bibinfo{author}{\bibfnamefont{H.}~\bibnamefont{Kjonsberg}},
  \bibinfo{author}{\bibfnamefont{A.~W.} \bibnamefont{Sandvik}},
  \bibnamefont{and} \bibinfo{author}{\bibfnamefont{S.~M.}
  \bibnamefont{Girvin}}, \bibinfo{journal}{Phys. Rev. B}
  \textbf{\bibinfo{volume}{70}}, \bibinfo{pages}{144411}
  (\bibinfo{year}{2004}).

\bibitem[{\citenamefont{Tabei et~al.}(2008{\natexlab{a}})\citenamefont{Tabei,
  Gingras, Kao, and Yavors'kii}}]{Tabei-MC}
\bibinfo{author}{\bibfnamefont{S.~M.~A.} \bibnamefont{Tabei}},
  \bibinfo{author}{\bibfnamefont{M.~J.~P.} \bibnamefont{Gingras}},
  \bibinfo{author}{\bibfnamefont{Y.-J.} \bibnamefont{Kao}}, \bibnamefont{and}
  \bibinfo{author}{\bibfnamefont{T.}~\bibnamefont{Yavors'kii}},
  \bibinfo{journal}{cond-mat/0801.0443}  (\bibinfo{year}{2008}{\natexlab{a}}).

\bibitem[{\citenamefont{Tabei et~al.}(2008{\natexlab{b}})\citenamefont{Tabei,
  Vernay, and Gingras}}]{TabeiVernay}
\bibinfo{author}{\bibfnamefont{S.~M.~A.} \bibnamefont{Tabei}},
  \bibinfo{author}{\bibfnamefont{F.}~\bibnamefont{Vernay}}, \bibnamefont{and}
  \bibinfo{author}{\bibfnamefont{M.~J.~P.} \bibnamefont{Gingras}},
  \bibinfo{journal}{Phys. Rev. B} \textbf{\bibinfo{volume}{77}},
  \bibinfo{pages}{014432} (\bibinfo{year}{2008}{\natexlab{b}}).

\bibitem[{\citenamefont{Binder and Young}(1986)}]{Binder}
\bibinfo{author}{\bibfnamefont{K.}~\bibnamefont{Binder}} \bibnamefont{and}
  \bibinfo{author}{\bibfnamefont{A.~P.} \bibnamefont{Young}},
  \bibinfo{journal}{Rev. Mod. Phys.} \textbf{\bibinfo{volume}{58}},
  \bibinfo{pages}{801} (\bibinfo{year}{1986}).

\bibitem[{\citenamefont{Mydosh}(1993)}]{Mydosh}
\bibinfo{author}{\bibfnamefont{J.~A.} \bibnamefont{Mydosh}},
  \emph{\bibinfo{title}{Spin Glasses: an Experimental Introduction}}
  (\bibinfo{publisher}{Taylor \& Francis, London}, \bibinfo{year}{1993}).

\bibitem[{\citenamefont{Rieger and Young}(1994)}]{Rieger-PRL}
\bibinfo{author}{\bibfnamefont{H.}~\bibnamefont{Rieger}} \bibnamefont{and}
  \bibinfo{author}{\bibfnamefont{A.~P.} \bibnamefont{Young}},
  \bibinfo{journal}{Phys. Rev. Lett.} \textbf{\bibinfo{volume}{72}},
  \bibinfo{pages}{4141} (\bibinfo{year}{1994}).

\bibitem[{\citenamefont{Rieger and Young}(1996)}]{Rieger-PRB}
\bibinfo{author}{\bibfnamefont{H.}~\bibnamefont{Rieger}} \bibnamefont{and}
  \bibinfo{author}{\bibfnamefont{A.~P.} \bibnamefont{Young}},
  \bibinfo{journal}{Phys. Rev. B} \textbf{\bibinfo{volume}{54}},
  \bibinfo{pages}{3328} (\bibinfo{year}{1996}).

\bibitem[{\citenamefont{Guo et~al.}(1996)\citenamefont{Guo, Bhatt, and
  Huse}}]{Guo}
\bibinfo{author}{\bibfnamefont{M.}~\bibnamefont{Guo}},
  \bibinfo{author}{\bibfnamefont{R.~N.} \bibnamefont{Bhatt}}, \bibnamefont{and}
  \bibinfo{author}{\bibfnamefont{D.~A.} \bibnamefont{Huse}},
  \bibinfo{journal}{Phys. Rev. B} \textbf{\bibinfo{volume}{54}},
  \bibinfo{pages}{3336} (\bibinfo{year}{1996}).

\bibitem[{\citenamefont{Griffiths}(1969)}]{Griffiths}
\bibinfo{author}{\bibfnamefont{R.~B.} \bibnamefont{Griffiths}},
  \bibinfo{journal}{Phys. Rev. Lett.} \textbf{\bibinfo{volume}{23}},
  \bibinfo{pages}{17} (\bibinfo{year}{1969}).

\bibitem[{\citenamefont{McCoy}(1969)}]{McCoY}
\bibinfo{author}{\bibfnamefont{B.~M.} \bibnamefont{McCoy}},
  \bibinfo{journal}{Phys. Rev. Lett.} \textbf{\bibinfo{volume}{23}},
  \bibinfo{pages}{383} (\bibinfo{year}{1969}).

\bibitem[{\citenamefont{Reich et~al.}(1990)\citenamefont{Reich, Ellman, Yang,
  Rosenbaum, Aeppli, and Belanger}}]{Reich}
\bibinfo{author}{\bibfnamefont{D.~H.} \bibnamefont{Reich}},
  \bibinfo{author}{\bibfnamefont{B.}~\bibnamefont{Ellman}},
  \bibinfo{author}{\bibfnamefont{J.}~\bibnamefont{Yang}},
  \bibinfo{author}{\bibfnamefont{T.~F.} \bibnamefont{Rosenbaum}},
  \bibinfo{author}{\bibfnamefont{G.}~\bibnamefont{Aeppli}}, \bibnamefont{and}
  \bibinfo{author}{\bibfnamefont{D.~P.} \bibnamefont{Belanger}},
  \bibinfo{journal}{Phys. Rev. B} \textbf{\bibinfo{volume}{42}},
  \bibinfo{pages}{4631} (\bibinfo{year}{1990}).

\bibitem[{\citenamefont{Wu et~al.}(1991)\citenamefont{Wu, Ellman, Rosenbaum,
  Aeppli, and Reich}}]{Wu-sg}
\bibinfo{author}{\bibfnamefont{W.}~\bibnamefont{Wu}},
  \bibinfo{author}{\bibfnamefont{B.}~\bibnamefont{Ellman}},
  \bibinfo{author}{\bibfnamefont{T.~F.} \bibnamefont{Rosenbaum}},
  \bibinfo{author}{\bibfnamefont{G.}~\bibnamefont{Aeppli}}, \bibnamefont{and}
  \bibinfo{author}{\bibfnamefont{D.~H.} \bibnamefont{Reich}},
  \bibinfo{journal}{Phys. Rev. Lett.} \textbf{\bibinfo{volume}{67}},
  \bibinfo{pages}{2076} (\bibinfo{year}{1991}).

\bibitem[{\citenamefont{Wu et~al.}(1993)\citenamefont{Wu, Bitko, Rosenbaum, and
  Aeppli}}]{Wu-chi3}
\bibinfo{author}{\bibfnamefont{W.}~\bibnamefont{Wu}},
  \bibinfo{author}{\bibfnamefont{D.}~\bibnamefont{Bitko}},
  \bibinfo{author}{\bibfnamefont{T.~F.} \bibnamefont{Rosenbaum}},
  \bibnamefont{and} \bibinfo{author}{\bibfnamefont{G.}~\bibnamefont{Aeppli}},
  \bibinfo{journal}{Phys. Rev. Lett.} \textbf{\bibinfo{volume}{71}},
  \bibinfo{pages}{1919} (\bibinfo{year}{1993}).

\bibitem[{\citenamefont{Wu}(1992)}]{Wu-thesis}
\bibinfo{author}{\bibfnamefont{W.}~\bibnamefont{Wu}}, \bibinfo{type}{{PhD}
  thesis}, \bibinfo{school}{U. of Chicago} (\bibinfo{year}{1992}).

\bibitem[{\citenamefont{Bitko et~al.}(1996)\citenamefont{Bitko, Rosenbaum, and
  Aeppli}}]{Bitko}
\bibinfo{author}{\bibfnamefont{D.}~\bibnamefont{Bitko}},
  \bibinfo{author}{\bibfnamefont{T.~F.} \bibnamefont{Rosenbaum}},
  \bibnamefont{and} \bibinfo{author}{\bibfnamefont{G.}~\bibnamefont{Aeppli}},
  \bibinfo{journal}{Phys. Rev. Lett.} \textbf{\bibinfo{volume}{77}},
  \bibinfo{pages}{940} (\bibinfo{year}{1996}).

\bibitem[{\citenamefont{Brooke et~al.}(1999)\citenamefont{Brooke, Bitko,
  Rosenbaum, and Aeppli}}]{Brooke}
\bibinfo{author}{\bibfnamefont{J.}~\bibnamefont{Brooke}},
  \bibinfo{author}{\bibfnamefont{D.}~\bibnamefont{Bitko}},
  \bibinfo{author}{\bibfnamefont{T.~F.} \bibnamefont{Rosenbaum}},
  \bibnamefont{and} \bibinfo{author}{\bibfnamefont{G.}~\bibnamefont{Aeppli}},
  \bibinfo{journal}{Science} \textbf{\bibinfo{volume}{284}},
  \bibinfo{pages}{779} (\bibinfo{year}{1999}).

\bibitem[{\citenamefont{Brooke}(2000)}]{Brooke-thesis}
\bibinfo{author}{\bibfnamefont{J.}~\bibnamefont{Brooke}}, Ph.D. thesis,
  \bibinfo{school}{U. of Chicago} (\bibinfo{year}{2000}).

\bibitem[{\citenamefont{Ghosh et~al.}(2002)\citenamefont{Ghosh, Parthasarathy,
  Rosenbaum, and Aeppli}}]{Ghosh-Science}
\bibinfo{author}{\bibfnamefont{S.}~\bibnamefont{Ghosh}},
  \bibinfo{author}{\bibfnamefont{R.}~\bibnamefont{Parthasarathy}},
  \bibinfo{author}{\bibfnamefont{T.~F.} \bibnamefont{Rosenbaum}},
  \bibnamefont{and} \bibinfo{author}{\bibfnamefont{G.}~\bibnamefont{Aeppli}},
  \bibinfo{journal}{Science} \textbf{\bibinfo{volume}{296}},
  \bibinfo{pages}{2195} (\bibinfo{year}{2002}).

\bibitem[{\citenamefont{Ghosh et~al.}(2003)\citenamefont{Ghosh, Rosenbaum,
  Aeppli, and Coppersmith}}]{Ghosh-Nature}
\bibinfo{author}{\bibfnamefont{S.}~\bibnamefont{Ghosh}},
  \bibinfo{author}{\bibfnamefont{T.~F.} \bibnamefont{Rosenbaum}},
  \bibinfo{author}{\bibfnamefont{G.}~\bibnamefont{Aeppli}}, \bibnamefont{and}
  \bibinfo{author}{\bibfnamefont{S.~N.} \bibnamefont{Coppersmith}},
  \bibinfo{journal}{Nature} \textbf{\bibinfo{volume}{425}}, \bibinfo{pages}{48}
  (\bibinfo{year}{2003}).

\bibitem[{\citenamefont{Ronnow et~al.}(2005)\citenamefont{Ronnow,
  Parthasarathy, Jensen, Aeppli, Rosenbaum, and McMorrow}}]{Ronnow}
\bibinfo{author}{\bibfnamefont{H.~M.} \bibnamefont{Ronnow}},
  \bibinfo{author}{\bibfnamefont{R.}~\bibnamefont{Parthasarathy}},
  \bibinfo{author}{\bibfnamefont{J.}~\bibnamefont{Jensen}},
  \bibinfo{author}{\bibfnamefont{G.}~\bibnamefont{Aeppli}},
  \bibinfo{author}{\bibfnamefont{T.~F.} \bibnamefont{Rosenbaum}},
  \bibnamefont{and} \bibinfo{author}{\bibfnamefont{D.~F.}
  \bibnamefont{McMorrow}}, \bibinfo{journal}{Science}
  \textbf{\bibinfo{volume}{308}}, \bibinfo{pages}{389} (\bibinfo{year}{2005}).

\bibitem[{\citenamefont{Ronnow et~al.}(2007)\citenamefont{Ronnow, Jensen,
  Parthasarathy, Aeppli, Rosenbaum, McMorrow, and Kraemer}}]{Ronnow2007}
\bibinfo{author}{\bibfnamefont{H.~M.} \bibnamefont{Ronnow}},
  \bibinfo{author}{\bibfnamefont{J.}~\bibnamefont{Jensen}},
  \bibinfo{author}{\bibfnamefont{R.}~\bibnamefont{Parthasarathy}},
  \bibinfo{author}{\bibfnamefont{G.}~\bibnamefont{Aeppli}},
  \bibinfo{author}{\bibfnamefont{T.~F.} \bibnamefont{Rosenbaum}},
  \bibinfo{author}{\bibfnamefont{D.~F.} \bibnamefont{McMorrow}},
  \bibnamefont{and} \bibinfo{author}{\bibfnamefont{C.}~\bibnamefont{Kraemer}},
  \bibinfo{journal}{Phys. Rev. B} \textbf{\bibinfo{volume}{75}},
  \bibinfo{pages}{054426} (\bibinfo{year}{2007}).

\bibitem[{\citenamefont{Silevitch
  et~al.}(2007{\natexlab{a}})\citenamefont{Silevitch, Bitko, Brooke, Ghosh,
  Aeppli, and Rosenbaum}}]{SilevitchNature}
\bibinfo{author}{\bibfnamefont{D.~M.} \bibnamefont{Silevitch}},
  \bibinfo{author}{\bibfnamefont{D.}~\bibnamefont{Bitko}},
  \bibinfo{author}{\bibfnamefont{J.}~\bibnamefont{Brooke}},
  \bibinfo{author}{\bibfnamefont{S.}~\bibnamefont{Ghosh}},
  \bibinfo{author}{\bibfnamefont{G.}~\bibnamefont{Aeppli}}, \bibnamefont{and}
  \bibinfo{author}{\bibfnamefont{T.~F.} \bibnamefont{Rosenbaum}},
  \bibinfo{journal}{Nature} \textbf{\bibinfo{volume}{448}},
  \bibinfo{pages}{567} (\bibinfo{year}{2007}{\natexlab{a}}).

\bibitem[{\citenamefont{Silevitch
  et~al.}(2007{\natexlab{b}})\citenamefont{Silevitch, Gannarelli, Fisher,
  Aeppli, and Rosenbaum}}]{SilevitchPRL}
\bibinfo{author}{\bibfnamefont{D.~M.} \bibnamefont{Silevitch}},
  \bibinfo{author}{\bibfnamefont{C.~M.~S.} \bibnamefont{Gannarelli}},
  \bibinfo{author}{\bibfnamefont{A.~J.} \bibnamefont{Fisher}},
  \bibinfo{author}{\bibfnamefont{G.}~\bibnamefont{Aeppli}}, \bibnamefont{and}
  \bibinfo{author}{\bibfnamefont{T.~F.} \bibnamefont{Rosenbaum}},
  \bibinfo{journal}{Phys. Rev. Lett} \textbf{\bibinfo{volume}{99}},
  \bibinfo{pages}{057203} (\bibinfo{year}{2007}{\natexlab{b}}).

\bibitem[{\citenamefont{J\"onsson et~al.}(2007)\citenamefont{J\"onsson,
  Mathieu, Wernsdorfer, Tkachuk, and Barbara}}]{Jonsson}
\bibinfo{author}{\bibfnamefont{P.~E.} \bibnamefont{J\"onsson}},
  \bibinfo{author}{\bibfnamefont{R.}~\bibnamefont{Mathieu}},
  \bibinfo{author}{\bibfnamefont{W.}~\bibnamefont{Wernsdorfer}},
  \bibinfo{author}{\bibfnamefont{A.~M.} \bibnamefont{Tkachuk}},
  \bibnamefont{and} \bibinfo{author}{\bibfnamefont{B.}~\bibnamefont{Barbara}},
  \bibinfo{journal}{Phys. Rev. Lett} \textbf{\bibinfo{volume}{98}},
  \bibinfo{pages}{256403} (\bibinfo{year}{2007}).

\bibitem[{\citenamefont{C.~Ancona-Torres and Rosenbaum}(2008)}]{SilevitchSG}
\bibinfo{author}{\bibfnamefont{G.~A.} \bibnamefont{C.~Ancona-Torres},
  \bibfnamefont{D.~M.~Silevitch}} \bibnamefont{and}
  \bibinfo{author}{\bibfnamefont{T.~F.} \bibnamefont{Rosenbaum}},
  \bibinfo{journal}{cond-mat/0801.2335}  (\bibinfo{year}{2008}).

\bibitem[{\citenamefont{J\"onsson et~al.}(2008)\citenamefont{J\"onsson,
  Mathieu, Wernsdorfer, Tkachuk, and Barbara}}]{Jonsson-new}
\bibinfo{author}{\bibfnamefont{P.~E.} \bibnamefont{J\"onsson}},
  \bibinfo{author}{\bibfnamefont{R.}~\bibnamefont{Mathieu}},
  \bibinfo{author}{\bibfnamefont{W.}~\bibnamefont{Wernsdorfer}},
  \bibinfo{author}{\bibfnamefont{A.~M.} \bibnamefont{Tkachuk}},
  \bibnamefont{and} \bibinfo{author}{\bibfnamefont{B.}~\bibnamefont{Barbara}},
  \bibinfo{journal}{cond-mat/0803.1357}  (\bibinfo{year}{2008}).

\bibitem[{rec()}]{recent}
\bibinfo{note}{A recent experimental study suggest that there might not even be
  a spin glass transition in LiHo$_x$Y$_{1-x}$F$_4$ for $x \le$ 0.16, even in
  zero $B_x$. See Ref.~[\onlinecite{Jonsson}] and the discussions in
  Refs.~[\onlinecite{SilevitchSG,Jonsson-new}].}

\bibitem[{\citenamefont{Schechter and Stamp}(2005)}]{Schechter-PRL1}
\bibinfo{author}{\bibfnamefont{M.}~\bibnamefont{Schechter}} \bibnamefont{and}
  \bibinfo{author}{\bibfnamefont{P.~C.~E.} \bibnamefont{Stamp}},
  \bibinfo{journal}{Phys. Rev. Lett.} \textbf{\bibinfo{volume}{95}},
  \bibinfo{pages}{267208} (\bibinfo{year}{2005}).

\bibitem[{\citenamefont{Schechter and Laflorencie}(2006)}]{Schechter-PRL2}
\bibinfo{author}{\bibfnamefont{M.}~\bibnamefont{Schechter}} \bibnamefont{and}
  \bibinfo{author}{\bibfnamefont{N.}~\bibnamefont{Laflorencie}},
  \bibinfo{journal}{Phys. Rev. Lett.} \textbf{\bibinfo{volume}{97}},
  \bibinfo{pages}{137204} (\bibinfo{year}{2006}).

\bibitem[{\citenamefont{Tabei et~al.}(2006)\citenamefont{Tabei, Gingras, Kao,
  Stasiak, and Fortin}}]{Tabei-PRL}
\bibinfo{author}{\bibfnamefont{S.~M.~A.} \bibnamefont{Tabei}},
  \bibinfo{author}{\bibfnamefont{M.~J.~P.} \bibnamefont{Gingras}},
  \bibinfo{author}{\bibfnamefont{Y.-J.} \bibnamefont{Kao}},
  \bibinfo{author}{\bibfnamefont{P.}~\bibnamefont{Stasiak}}, \bibnamefont{and}
  \bibinfo{author}{\bibfnamefont{J.-Y.} \bibnamefont{Fortin}},
  \bibinfo{journal}{Phys. Rev. Lett.} \textbf{\bibinfo{volume}{97}},
  \bibinfo{pages}{237203} (\bibinfo{year}{2006}).

\bibitem[{\citenamefont{Quilliam et~al.}(2007)\citenamefont{Quilliam, Mugford,
  Gomez, Kycia, and Kycia}}]{Quilliam}
\bibinfo{author}{\bibfnamefont{J.~A.} \bibnamefont{Quilliam}},
  \bibinfo{author}{\bibfnamefont{C.~G.~A.} \bibnamefont{Mugford}},
  \bibinfo{author}{\bibfnamefont{A.}~\bibnamefont{Gomez}},
  \bibinfo{author}{\bibfnamefont{S.~W.} \bibnamefont{Kycia}}, \bibnamefont{and}
  \bibinfo{author}{\bibfnamefont{J.~B.} \bibnamefont{Kycia}},
  \bibinfo{journal}{Phys. Rev. Lett} \textbf{\bibinfo{volume}{98}},
  \bibinfo{pages}{037203} (\bibinfo{year}{2007}).

\bibitem[{\citenamefont{Stephen and Aharony}(1981)}]{Stephen}
\bibinfo{author}{\bibfnamefont{M.~J.} \bibnamefont{Stephen}} \bibnamefont{and}
  \bibinfo{author}{\bibfnamefont{A.}~\bibnamefont{Aharony}},
  \bibinfo{journal}{J. Phys. C} \textbf{\bibinfo{volume}{14}},
  \bibinfo{pages}{1665} (\bibinfo{year}{1981}).

\bibitem[{\citenamefont{Snyder and Yu}(2005)}]{Snyder}
\bibinfo{author}{\bibfnamefont{J.}~\bibnamefont{Snyder}} \bibnamefont{and}
  \bibinfo{author}{\bibfnamefont{C.~C.} \bibnamefont{Yu}},
  \bibinfo{journal}{Phys. Rev. B} \textbf{\bibinfo{volume}{72}},
  \bibinfo{pages}{214203} (\bibinfo{year}{2005}).

\bibitem[{\citenamefont{Biltmo and Henelius}(2007)}]{Biltmo1}
\bibinfo{author}{\bibfnamefont{A.}~\bibnamefont{Biltmo}} \bibnamefont{and}
  \bibinfo{author}{\bibfnamefont{P.}~\bibnamefont{Henelius}},
  \bibinfo{journal}{Phys. Rev. B} \textbf{\bibinfo{volume}{76}},
  \bibinfo{pages}{054423} (\bibinfo{year}{2007}).

\bibitem[{\citenamefont{Biltmo and Henelius}(2008)}]{Biltmo2}
\bibinfo{author}{\bibfnamefont{A.}~\bibnamefont{Biltmo}} \bibnamefont{and}
  \bibinfo{author}{\bibfnamefont{P.}~\bibnamefont{Henelius}},
  \bibinfo{journal}{arXiv:0803.0851}  (\bibinfo{year}{2008}).

\bibitem[{\citenamefont{Bhatt and Lee}(1982)}]{Bhatt}
\bibinfo{author}{\bibfnamefont{R.~N.} \bibnamefont{Bhatt}} \bibnamefont{and}
  \bibinfo{author}{\bibfnamefont{P.~A.} \bibnamefont{Lee}},
  \bibinfo{journal}{Phys. Rev. Lett.} \textbf{\bibinfo{volume}{48}},
  \bibinfo{pages}{344} (\bibinfo{year}{1982}).

\bibitem[{\citenamefont{Chin and Eastham}(2006)}]{Chin}
\bibinfo{author}{\bibfnamefont{A.}~\bibnamefont{Chin}} \bibnamefont{and}
  \bibinfo{author}{\bibfnamefont{P.~R.} \bibnamefont{Eastham}},
  \bibinfo{journal}{cond-mat/0610544}  (\bibinfo{year}{2006}).

\bibitem[{Tam()}]{Tam}
\bibinfo{note}{K.-M. Tam and M. J.P. Gingras (unpublished).}

\bibitem[{\citenamefont{Shakurov et~al.}(2005)\citenamefont{Shakurov, Vanyunin,
  Malkin, Barbara, Abdulsabirov, and Korableva}}]{Sharukov}
\bibinfo{author}{\bibfnamefont{G.~S.} \bibnamefont{Shakurov}},
  \bibinfo{author}{\bibfnamefont{M.~V.} \bibnamefont{Vanyunin}},
  \bibinfo{author}{\bibfnamefont{B.~Z.} \bibnamefont{Malkin}},
  \bibinfo{author}{\bibfnamefont{B.}~\bibnamefont{Barbara}},
  \bibinfo{author}{\bibfnamefont{R.~Y.} \bibnamefont{Abdulsabirov}},
  \bibnamefont{and} \bibinfo{author}{\bibfnamefont{S.~L.}
  \bibnamefont{Korableva}}, \bibinfo{journal}{Appl. Magn. Reson.}
  \textbf{\bibinfo{volume}{28}}, \bibinfo{pages}{251} (\bibinfo{year}{2005}).

\bibitem[{\citenamefont{Bertaina et~al.}(2006)\citenamefont{Bertaina, Barbara,
  Giraud, Malkin, Vanuynin, Pominov, Stolov, and Tkachuk}}]{Bertaina}
\bibinfo{author}{\bibfnamefont{S.}~\bibnamefont{Bertaina}},
  \bibinfo{author}{\bibfnamefont{B.}~\bibnamefont{Barbara}},
  \bibinfo{author}{\bibfnamefont{R.}~\bibnamefont{Giraud}},
  \bibinfo{author}{\bibfnamefont{B.~Z.} \bibnamefont{Malkin}},
  \bibinfo{author}{\bibfnamefont{M.~V.} \bibnamefont{Vanuynin}},
  \bibinfo{author}{\bibfnamefont{A.~I.} \bibnamefont{Pominov}},
  \bibinfo{author}{\bibfnamefont{A.~L.} \bibnamefont{Stolov}},
  \bibnamefont{and} \bibinfo{author}{\bibfnamefont{A.~M.}
  \bibnamefont{Tkachuk}}, \bibinfo{journal}{Phys. Rev. B}
  \textbf{\bibinfo{volume}{74}}, \bibinfo{pages}{184421}
  (\bibinfo{year}{2006}).

\bibitem[{\citenamefont{Agladze et~al.}(1991)\citenamefont{Agladze, Popova,
  Zhizhin, Egorov, and Petrova}}]{Agladze}
\bibinfo{author}{\bibfnamefont{N.~I.} \bibnamefont{Agladze}},
  \bibinfo{author}{\bibfnamefont{M.~N.} \bibnamefont{Popova}},
  \bibinfo{author}{\bibfnamefont{G.~N.} \bibnamefont{Zhizhin}},
  \bibinfo{author}{\bibfnamefont{V.~J.} \bibnamefont{Egorov}},
  \bibnamefont{and} \bibinfo{author}{\bibfnamefont{M.~A.}
  \bibnamefont{Petrova}}, \bibinfo{journal}{Phys. Rev. Lett.}
  \textbf{\bibinfo{volume}{66}}, \bibinfo{pages}{477} (\bibinfo{year}{1991}).

\bibitem[{\citenamefont{Fortin and Gingras}(2007)}]{Fortin}
\bibinfo{author}{\bibfnamefont{J.-Y.} \bibnamefont{Fortin}} \bibnamefont{and}
  \bibinfo{author}{\bibfnamefont{M.~J.~P.} \bibnamefont{Gingras}},
  \bibinfo{journal}{unpublished}  (\bibinfo{year}{2007}).

\bibitem[{\citenamefont{Magarino et~al.}(1980)\citenamefont{Magarino,
  Tuchendler, Beauvillain, and Laursen}}]{Magarino}
\bibinfo{author}{\bibfnamefont{J.}~\bibnamefont{Magarino}},
  \bibinfo{author}{\bibfnamefont{J.}~\bibnamefont{Tuchendler}},
  \bibinfo{author}{\bibfnamefont{P.}~\bibnamefont{Beauvillain}},
  \bibnamefont{and} \bibinfo{author}{\bibfnamefont{I.}~\bibnamefont{Laursen}},
  \bibinfo{journal}{Phys. Rev. B} \textbf{\bibinfo{volume}{21}},
  \bibinfo{pages}{18} (\bibinfo{year}{1980}).

\bibitem[{\citenamefont{Kazei et~al.}(2006)\citenamefont{Kazei, Snegirev,
  Chanieva, Abdulsabirov, and Korableva}}]{Kazei}
\bibinfo{author}{\bibfnamefont{Z.~A.} \bibnamefont{Kazei}},
  \bibinfo{author}{\bibfnamefont{V.~V.} \bibnamefont{Snegirev}},
  \bibinfo{author}{\bibfnamefont{R.~I.} \bibnamefont{Chanieva}},
  \bibinfo{author}{\bibfnamefont{R.~Y.} \bibnamefont{Abdulsabirov}},
  \bibnamefont{and} \bibinfo{author}{\bibfnamefont{S.~L.}
  \bibnamefont{Korableva}}, \bibinfo{journal}{Phys. of the Solid State}
  \textbf{\bibinfo{volume}{48}}, \bibinfo{pages}{726} (\bibinfo{year}{2006}).

\bibitem[{\citenamefont{Liu et~al.}(1988)\citenamefont{Liu, Huang, Cone, and
  Jacquier}}]{LiTbF4-singlet}
\bibinfo{author}{\bibfnamefont{G.~K.} \bibnamefont{Liu}},
  \bibinfo{author}{\bibfnamefont{J.}~\bibnamefont{Huang}},
  \bibinfo{author}{\bibfnamefont{R.~L.} \bibnamefont{Cone}}, \bibnamefont{and}
  \bibinfo{author}{\bibfnamefont{B.}~\bibnamefont{Jacquier}},
  \bibinfo{journal}{Phys. Rev. B} \textbf{\bibinfo{volume}{38}},
  \bibinfo{pages}{11061} (\bibinfo{year}{1988}).

\bibitem[{\citenamefont{Aeppli}(1998)}]{Aeppli-Nato}
\bibinfo{author}{\bibfnamefont{G.}~\bibnamefont{Aeppli}}, in
  \emph{\bibinfo{booktitle}{Proceedings of the NATO Advanced Study Institute on
  Dynamical Properties of Unconventional Magnetic Systems}}, edited by
  \bibinfo{editor}{\bibfnamefont{A.~T.} \bibnamefont{Skjeltorp}}
  \bibnamefont{and}
  \bibinfo{editor}{\bibfnamefont{D.}~\bibnamefont{Sherrington}}
  (\bibinfo{publisher}{Kluwer Academic Publishers}, \bibinfo{year}{1998}).

\bibitem[{spi()}]{spin-ice}
\bibinfo{note}{In recent years, the frustrated ferromagnetic spin ice materials
  have attracted much interest~\cite{Bramwell-SI}. These systems are extremely
  well described by a classical Ising model where the leading interactions are
  long-range magnetic dipole-dipole couplings in addition to a short-range
  exchange interaction~\cite{Hertog,Yavorskii-SI}. The application of a
  magnetic field on these systems have been found to give rise to interesting
  effects~\cite{Fennell-SI-field}. There has also been studies of diluted spin
  ice compounds~\cite{Ke}. However, the excited crystal field levels of so far
  considered Dy$_2$Ti$_2$O$_7$ and Ho$_2$Ti$_2$O$_7$ spin ice materials are at
  very high energy compared to typically accessible magnetic field Zeeman
  energy~\cite{Rosenkranz}. As a result, field-induced quantum mechanical
  fluctuation effects are negligible in spin ice materials~\cite{Ruff-field}
  and these are unfortunately not suitable materials for the purpose of
  exploring TFIM physics. One exception may be Tb$_2$Ti$_2$O$_7$ where, because
  of the smallest energy gap to the lowest excited
  doublet~\cite{Rosenkranz,Gingras-PRB2000,Mirebeau-INS} field-induced quantum
  effects may perhaps be investigated~\cite{Rule}.}

\bibitem[{\citenamefont{Catanese et~al.}(1973)\citenamefont{Catanese,
  Skjeltorp, Meissner, and Wolf}}]{Catanese-Tb}
\bibinfo{author}{\bibfnamefont{C.~A.} \bibnamefont{Catanese}},
  \bibinfo{author}{\bibfnamefont{A.~T.} \bibnamefont{Skjeltorp}},
  \bibinfo{author}{\bibfnamefont{H.~E.} \bibnamefont{Meissner}},
  \bibnamefont{and} \bibinfo{author}{\bibfnamefont{W.~P.} \bibnamefont{Wolf}},
  \bibinfo{journal}{Phys. Rev. B} \textbf{\bibinfo{volume}{8}},
  \bibinfo{pages}{4223} (\bibinfo{year}{1973}).

\bibitem[{\citenamefont{Scott}(1970)}]{Scott-thesis}
\bibinfo{author}{\bibfnamefont{P.~D.} \bibnamefont{Scott}}, Ph.D. thesis,
  \bibinfo{school}{Yale University} (\bibinfo{year}{1970}).

\bibitem[{\citenamefont{Beall et~al.}(1977)\citenamefont{Beall, Milligan, and
  Wolcott}}]{Beall}
\bibinfo{author}{\bibfnamefont{G.~W.} \bibnamefont{Beall}},
  \bibinfo{author}{\bibfnamefont{W.~O.} \bibnamefont{Milligan}},
  \bibnamefont{and} \bibinfo{author}{\bibfnamefont{H.~A.}
  \bibnamefont{Wolcott}}, \bibinfo{journal}{J. Inorg. Nuc. Chem.}
  \textbf{\bibinfo{volume}{39}}, \bibinfo{pages}{65} (\bibinfo{year}{1977}).

\bibitem[{\citenamefont{Landers and Brun}(1973)}]{Landers}
\bibinfo{author}{\bibfnamefont{G.~H.} \bibnamefont{Landers}} \bibnamefont{and}
  \bibinfo{author}{\bibfnamefont{T.~O.} \bibnamefont{Brun}},
  \bibinfo{journal}{Acta Cryst.} \textbf{\bibinfo{volume}{A29}},
  \bibinfo{pages}{684} (\bibinfo{year}{1973}).

\bibitem[{\citenamefont{Scott et~al.}(1969)\citenamefont{Scott, Meissner, and
  Crosswhite}}]{Scott-Tb}
\bibinfo{author}{\bibfnamefont{P.~D.} \bibnamefont{Scott}},
  \bibinfo{author}{\bibfnamefont{H.~E.} \bibnamefont{Meissner}},
  \bibnamefont{and} \bibinfo{author}{\bibfnamefont{H.~M.}
  \bibnamefont{Crosswhite}}, \bibinfo{journal}{Phys. Lett. A}
  \textbf{\bibinfo{volume}{28}}, \bibinfo{pages}{489} (\bibinfo{year}{1969}).

\bibitem[{\citenamefont{Kahle et~al.}(1986)\citenamefont{Kahle, Kasten, Scott,
  and Wolf}}]{Scott-Dy}
\bibinfo{author}{\bibfnamefont{H.~G.} \bibnamefont{Kahle}},
  \bibinfo{author}{\bibfnamefont{A.}~\bibnamefont{Kasten}},
  \bibinfo{author}{\bibfnamefont{P.~D.} \bibnamefont{Scott}}, \bibnamefont{and}
  \bibinfo{author}{\bibfnamefont{W.~P.} \bibnamefont{Wolf}},
  \bibinfo{journal}{J. Phys. C} \textbf{\bibinfo{volume}{19}},
  \bibinfo{pages}{4153} (\bibinfo{year}{1986}).

\bibitem[{\citenamefont{Stevens}(1952)}]{Stevens}
\bibinfo{author}{\bibfnamefont{K.~W.~H.} \bibnamefont{Stevens}},
  \bibinfo{journal}{Proc. Phys. Soc. A} \textbf{\bibinfo{volume}{65}},
  \bibinfo{pages}{209} (\bibinfo{year}{1952}).

\bibitem[{\citenamefont{Hutchings}(1964)}]{Hutchings}
\bibinfo{author}{\bibfnamefont{M.~T.} \bibnamefont{Hutchings}},
  \emph{\bibinfo{title}{Point-Charge Calculations of Energy Levels of Magnetic
  Ions in Crystaline Electric Fields}} (\bibinfo{publisher}{Academic Press},
  \bibinfo{address}{New York and London}, \bibinfo{year}{1964}),
  vol.~\bibinfo{volume}{16} of \emph{\bibinfo{series}{Solid State Physics}}, p.
  \bibinfo{pages}{227}.

\bibitem[{\citenamefont{Karmakar et~al.}(1981)\citenamefont{Karmakar, Saha, and
  Ghosh}}]{Karmakar-Ho}
\bibinfo{author}{\bibfnamefont{S.}~\bibnamefont{Karmakar}},
  \bibinfo{author}{\bibfnamefont{M.}~\bibnamefont{Saha}}, \bibnamefont{and}
  \bibinfo{author}{\bibfnamefont{D.}~\bibnamefont{Ghosh}}, \bibinfo{journal}{J.
  App. Phys.} \textbf{\bibinfo{volume}{52}}, \bibinfo{pages}{4156}
  (\bibinfo{year}{1981}).

\bibitem[{\citenamefont{Karmakar et~al.}(1982)\citenamefont{Karmakar, Saha, and
  Ghosh}}]{Karmakar-Dy}
\bibinfo{author}{\bibfnamefont{S.}~\bibnamefont{Karmakar}},
  \bibinfo{author}{\bibfnamefont{M.}~\bibnamefont{Saha}}, \bibnamefont{and}
  \bibinfo{author}{\bibfnamefont{D.}~\bibnamefont{Ghosh}},
  \bibinfo{journal}{Phys. Rev. B} \textbf{\bibinfo{volume}{26}},
  \bibinfo{pages}{7023} (\bibinfo{year}{1982}).

\bibitem[{Arb()}]{Arbitrary}
\bibinfo{note}{By making this choice, we do not imply higher validity or better
  acurracy of the chosen sets of crystal field parameters.}

\bibitem[{\citenamefont{Scott and Wolf}(1969)}]{Scott-Tb-2}
\bibinfo{author}{\bibfnamefont{P.~D.} \bibnamefont{Scott}} \bibnamefont{and}
  \bibinfo{author}{\bibfnamefont{W.~P.} \bibnamefont{Wolf}},
  \bibinfo{journal}{J. App. Phys.} \textbf{\bibinfo{volume}{40}},
  \bibinfo{pages}{1031} (\bibinfo{year}{1969}).

\bibitem[{\citenamefont{Catanese and Meissner}(1973)}]{Catanese-DyHo}
\bibinfo{author}{\bibfnamefont{C.~A.} \bibnamefont{Catanese}} \bibnamefont{and}
  \bibinfo{author}{\bibfnamefont{H.~E.} \bibnamefont{Meissner}},
  \bibinfo{journal}{Phys. Rev. B} \textbf{\bibinfo{volume}{8}},
  \bibinfo{pages}{2060} (\bibinfo{year}{1973}).

\bibitem[{\citenamefont{Ewald}(1921)}]{Ewald}
\bibinfo{author}{\bibfnamefont{P.~P.} \bibnamefont{Ewald}},
  \bibinfo{journal}{Annalen der Physik} \textbf{\bibinfo{volume}{64}},
  \bibinfo{pages}{253} (\bibinfo{year}{1921}).

\bibitem[{\citenamefont{Ziman}(1972)}]{Ziman}
\bibinfo{author}{\bibfnamefont{J.~M.} \bibnamefont{Ziman}},
  \emph{\bibinfo{title}{Principles of the Theory of Solids}}
  (\bibinfo{publisher}{Cambridge University Press}, \bibinfo{year}{1972}).

\bibitem[{\citenamefont{de~Leeuw et~al.}(1980)\citenamefont{de~Leeuw, Perram,
  and Smith}}]{Leeuw}
\bibinfo{author}{\bibfnamefont{S.~W.} \bibnamefont{de~Leeuw}},
  \bibinfo{author}{\bibfnamefont{J.~W.} \bibnamefont{Perram}},
  \bibnamefont{and} \bibinfo{author}{\bibfnamefont{E.~R.} \bibnamefont{Smith}},
  \bibinfo{journal}{Proc. Soc. Lond. A} \textbf{\bibinfo{volume}{373}},
  \bibinfo{pages}{27} (\bibinfo{year}{1980}).

\bibitem[{\citenamefont{Born and Huang}(1968)}]{Born}
\bibinfo{author}{\bibfnamefont{M.}~\bibnamefont{Born}} \bibnamefont{and}
  \bibinfo{author}{\bibfnamefont{K.}~\bibnamefont{Huang}},
  \emph{\bibinfo{title}{Theory of Crystal Lattices}}
  (\bibinfo{publisher}{Oxford University Press, London}, \bibinfo{year}{1968}).

\bibitem[{\citenamefont{Melko and Gingras}(2004)}]{Melko-JPC}
\bibinfo{author}{\bibfnamefont{R.~G.} \bibnamefont{Melko}} \bibnamefont{and}
  \bibinfo{author}{\bibfnamefont{M.~J.~P.} \bibnamefont{Gingras}},
  \bibinfo{journal}{J. Phys.: Condens. Matter} \textbf{\bibinfo{volume}{16}},
  \bibinfo{pages}{R1277} (\bibinfo{year}{2004}).

\bibitem[{\citenamefont{Catanese}(1970)}]{Catanese-thesis}
\bibinfo{author}{\bibfnamefont{C.~A.} \bibnamefont{Catanese}}, Ph.D. thesis,
  \bibinfo{school}{Yale University} (\bibinfo{year}{1970}).

\bibitem[{\citenamefont{Tigges and Wolf}(1987)}]{Wolf-Tigges}
\bibinfo{author}{\bibfnamefont{C.~P.} \bibnamefont{Tigges}} \bibnamefont{and}
  \bibinfo{author}{\bibfnamefont{W.~P.} \bibnamefont{Wolf}},
  \bibinfo{journal}{Phys. Rev. Lett.} \textbf{\bibinfo{volume}{58}},
  \bibinfo{pages}{2371} (\bibinfo{year}{1987}).

\bibitem[{\citenamefont{Wolf}(2000)}]{Wolf}
\bibinfo{author}{\bibfnamefont{W.~P.} \bibnamefont{Wolf}},
  \bibinfo{journal}{Bras. J. Phys.} \textbf{\bibinfo{volume}{30}},
  \bibinfo{pages}{794} (\bibinfo{year}{2000}).

\bibitem[{\citenamefont{Jensen and Mackintosh}(1991)}]{Jensen}
\bibinfo{author}{\bibfnamefont{J.}~\bibnamefont{Jensen}} \bibnamefont{and}
  \bibinfo{author}{\bibfnamefont{A.~R.} \bibnamefont{Mackintosh}},
  \emph{\bibinfo{title}{Rare Earth Magnetism Theory $-$ Structures and
  Excitations}} (\bibinfo{publisher}{Clarendon Press, Oxford},
  \bibinfo{year}{1991}).

\bibitem[{ref()}]{referee}
\bibinfo{note}{We thank an anonymous referee for his/her comments on this
  point.}

\bibitem[{\citenamefont{K.~Deguchi and Maeno}(2004)}]{magnet}
\bibinfo{author}{\bibfnamefont{T.~I.} \bibnamefont{K.~Deguchi}}
  \bibnamefont{and} \bibinfo{author}{\bibfnamefont{Y.}~\bibnamefont{Maeno}},
  \bibinfo{journal}{Rev. Sci. Instrum.} \textbf{\bibinfo{volume}{75}},
  \bibinfo{pages}{1188} (\bibinfo{year}{2004}).

\bibitem[{\citenamefont{Bramwell and Gingras}(2001)}]{Bramwell-SI}
\bibinfo{author}{\bibfnamefont{S.~T.} \bibnamefont{Bramwell}} \bibnamefont{and}
  \bibinfo{author}{\bibfnamefont{M.~J.~P.} \bibnamefont{Gingras}},
  \bibinfo{journal}{Science} \textbf{\bibinfo{volume}{294}},
  \bibinfo{pages}{495} (\bibinfo{year}{2001}).

\bibitem[{\citenamefont{den Hertog and Gingras}(2000)}]{Hertog}
\bibinfo{author}{\bibfnamefont{B.~C.} \bibnamefont{den Hertog}}
  \bibnamefont{and} \bibinfo{author}{\bibfnamefont{M.~J.~P.}
  \bibnamefont{Gingras}}, \bibinfo{journal}{Phys. Rev. Lett.}
  \textbf{\bibinfo{volume}{84}}, \bibinfo{pages}{3430} (\bibinfo{year}{2000}).

\bibitem[{\citenamefont{Yavors'kii et~al.}(2007)\citenamefont{Yavors'kii,
  Fennell, Gingras, and Bramwell}}]{Yavorskii-SI}
\bibinfo{author}{\bibfnamefont{T.}~\bibnamefont{Yavors'kii}},
  \bibinfo{author}{\bibfnamefont{T.}~\bibnamefont{Fennell}},
  \bibinfo{author}{\bibfnamefont{M.~J.~P.} \bibnamefont{Gingras}},
  \bibnamefont{and} \bibinfo{author}{\bibfnamefont{S.~T.}
  \bibnamefont{Bramwell}}, \bibinfo{journal}{arXiv:0707.3477}
  (\bibinfo{year}{2007}).

\bibitem[{\citenamefont{Fennell et~al.}(2005)\citenamefont{Fennell, Petrenko,
  k, Gardner, Bramwell, and Ouladdiaf}}]{Fennell-SI-field}
\bibinfo{author}{\bibfnamefont{T.}~\bibnamefont{Fennell}},
  \bibinfo{author}{\bibfnamefont{O.~A.} \bibnamefont{Petrenko}},
  \bibinfo{author}{\bibfnamefont{B.~F.} \bibnamefont{k}},
  \bibinfo{author}{\bibfnamefont{J.~S.} \bibnamefont{Gardner}},
  \bibinfo{author}{\bibfnamefont{S.~T.} \bibnamefont{Bramwell}},
  \bibnamefont{and}
  \bibinfo{author}{\bibfnamefont{B.}~\bibnamefont{Ouladdiaf}},
  \bibinfo{journal}{Phys. Rev.B} \textbf{\bibinfo{volume}{72}},
  \bibinfo{pages}{224411} (\bibinfo{year}{2005}).

\bibitem[{\citenamefont{Ke et~al.}(2007)\citenamefont{Ke, Freitas, Ueland, Lau,
  Dahlberg, Cava, Moessner, and Schiffer}}]{Ke}
\bibinfo{author}{\bibfnamefont{X.}~\bibnamefont{Ke}},
  \bibinfo{author}{\bibfnamefont{R.~S.} \bibnamefont{Freitas}},
  \bibinfo{author}{\bibfnamefont{B.~G.} \bibnamefont{Ueland}},
  \bibinfo{author}{\bibfnamefont{G.~C.} \bibnamefont{Lau}},
  \bibinfo{author}{\bibfnamefont{M.~L.} \bibnamefont{Dahlberg}},
  \bibinfo{author}{\bibfnamefont{R.~J.} \bibnamefont{Cava}},
  \bibinfo{author}{\bibfnamefont{R.}~\bibnamefont{Moessner}}, \bibnamefont{and}
  \bibinfo{author}{\bibfnamefont{P.}~\bibnamefont{Schiffer}},
  \bibinfo{journal}{Phys. Rev. Lett.} \textbf{\bibinfo{volume}{99}},
  \bibinfo{pages}{137203} (\bibinfo{year}{2007}).

\bibitem[{\citenamefont{Rosenkranz et~al.}(2000)\citenamefont{Rosenkranz,
  Ramirez, Hayashi, Cava, Siddharthan, and Shastry}}]{Rosenkranz}
\bibinfo{author}{\bibfnamefont{S.}~\bibnamefont{Rosenkranz}},
  \bibinfo{author}{\bibfnamefont{A.~P.} \bibnamefont{Ramirez}},
  \bibinfo{author}{\bibfnamefont{A.}~\bibnamefont{Hayashi}},
  \bibinfo{author}{\bibfnamefont{R.~J.} \bibnamefont{Cava}},
  \bibinfo{author}{\bibfnamefont{R.}~\bibnamefont{Siddharthan}},
  \bibnamefont{and} \bibinfo{author}{\bibfnamefont{B.~S.}
  \bibnamefont{Shastry}}, \bibinfo{journal}{J. Appl. Phys.}
  \textbf{\bibinfo{volume}{87}}, \bibinfo{pages}{5914} (\bibinfo{year}{2000}).

\bibitem[{\citenamefont{Ruff et~al.}(2005)\citenamefont{Ruff, Melko, and
  Gingras}}]{Ruff-field}
\bibinfo{author}{\bibfnamefont{J.~P.~C.} \bibnamefont{Ruff}},
  \bibinfo{author}{\bibfnamefont{R.~G.} \bibnamefont{Melko}}, \bibnamefont{and}
  \bibinfo{author}{\bibfnamefont{M.~J.~P.} \bibnamefont{Gingras}},
  \bibinfo{journal}{Phys. Rev. Lett.} \textbf{\bibinfo{volume}{95}},
  \bibinfo{pages}{097202} (\bibinfo{year}{2005}).

\bibitem[{\citenamefont{Gingras et~al.}(2000)\citenamefont{Gingras, den Hertog,
  Faucher, Gardner, Chang, Gaulin, Raju, and Greedan}}]{Gingras-PRB2000}
\bibinfo{author}{\bibfnamefont{M.~J.~P.} \bibnamefont{Gingras}},
  \bibinfo{author}{\bibfnamefont{B.~C.} \bibnamefont{den Hertog}},
  \bibinfo{author}{\bibfnamefont{M.}~\bibnamefont{Faucher}},
  \bibinfo{author}{\bibfnamefont{J.}~\bibnamefont{Gardner}},
  \bibinfo{author}{\bibfnamefont{L.}~\bibnamefont{Chang}},
  \bibinfo{author}{\bibfnamefont{B.}~\bibnamefont{Gaulin}},
  \bibinfo{author}{\bibfnamefont{N.}~\bibnamefont{Raju}}, \bibnamefont{and}
  \bibinfo{author}{\bibfnamefont{J.}~\bibnamefont{Greedan}},
  \bibinfo{journal}{Phys. Rev. B} \textbf{\bibinfo{volume}{62}},
  \bibinfo{pages}{6496} (\bibinfo{year}{2000}).

\bibitem[{\citenamefont{Mirebeau et~al.}(2007)\citenamefont{Mirebeau, Bonville,
  and Hennion}}]{Mirebeau-INS}
\bibinfo{author}{\bibfnamefont{I.}~\bibnamefont{Mirebeau}},
  \bibinfo{author}{\bibfnamefont{P.}~\bibnamefont{Bonville}}, \bibnamefont{and}
  \bibinfo{author}{\bibfnamefont{M.}~\bibnamefont{Hennion}},
  \bibinfo{journal}{Phys. Rev. B} \textbf{\bibinfo{volume}{76}},
  \bibinfo{pages}{184436} (\bibinfo{year}{2007}).

\bibitem[{\citenamefont{Rule et~al.}(2006)\citenamefont{Rule, Ruff, Gaulin,
  Dunsiger, Gardner, Clancy, Lewis, Dabkowska, Mirebeau, Manuel et~al.}}]{Rule}
\bibinfo{author}{\bibfnamefont{K.~C.} \bibnamefont{Rule}},
  \bibinfo{author}{\bibfnamefont{J.~P.~C.} \bibnamefont{Ruff}},
  \bibinfo{author}{\bibfnamefont{B.~D.} \bibnamefont{Gaulin}},
  \bibinfo{author}{\bibfnamefont{S.~R.} \bibnamefont{Dunsiger}},
  \bibinfo{author}{\bibfnamefont{J.~S.} \bibnamefont{Gardner}},
  \bibinfo{author}{\bibfnamefont{J.~P.} \bibnamefont{Clancy}},
  \bibinfo{author}{\bibfnamefont{M.~J.} \bibnamefont{Lewis}},
  \bibinfo{author}{\bibfnamefont{H.~A.} \bibnamefont{Dabkowska}},
  \bibinfo{author}{\bibfnamefont{I.}~\bibnamefont{Mirebeau}},
  \bibinfo{author}{\bibfnamefont{P.}~\bibnamefont{Manuel}},
  \bibnamefont{et~al.}, \bibinfo{journal}{Phys. Rev. Lett.}
  \textbf{\bibinfo{volume}{96}}, \bibinfo{pages}{177201}
  (\bibinfo{year}{2006}).

\end{thebibliography}


\end{document}